\documentclass[useAMS,usenatbib]{mn2e}

\usepackage{graphicx}
\usepackage{txfonts}
\usepackage[breaklinks=true]{hyperref}
\usepackage{longtable}
\usepackage{subfig}
\usepackage{float}
\usepackage{booktabs}
\usepackage{natbib}
\usepackage{ amssymb }
\newcommand{\gsim}{\lower.7ex\hbox{$\;\stackrel{\textstyle>}{\sim}\;$}}

\def\aj{AJ}%
\def\araa{ARA\&A}%
\def\apj{ApJ}%
\def\apjl{ApJ}%
\def\apjs{ApJS}%
\def\aap{A\&A}%
\def\aaps{A\&AS}%
\def\mnras{MNRAS}%
\def\pasp{PASP}%
\def\rmxaa{Rev. Mexicana Astron. Astrofis.}%
\def\ssr{Space~Sci.~Rev.}%
\def\nat{Nature}%
\def\procspie{Proc.~SPIE}%

\title[AGN in Mrk\,1498]{A young and obscured AGN embedded in the giant radio galaxy Mrk\,1498}
\author[Hern\'{a}ndez-Garc\'{i}a et al.]{L. Hern\'{a}ndez-Garc\'{i}a$^{1}$\thanks{E-mail:
lorena.hernandez@uv.cl}, F. Panessa$^{2}$, L. Bassani$^{3}$, G. Bruni$^{2}$, F. Ursini$^{3}$, V. Chavushyan$^{4}$, 
\newauthor
O. Gonz\'alez-Mart\'in$^{5}$, S. Cazzoli$^{6}$, E. F. Jim\'enez-Andrade$^{7,8}$, P. Ar\'evalo$^{1}$, Y. D\'iaz$^{1}$, 
\newauthor
A. Bazzano$^{2}$, and P. Ubertini$^{2}$\\
$^{1}$Instituto de F\'isica y Astronom\'ia, Facultad de Ciencias, Universidad de Valpara\'iso, Gran Breta\~na 1111, Playa Ancha, Valpara\'iso, Chile \\
$^{2}$INAF - Istituto di Astrofisica e Planetologia Spaziali (IAPS-INAF), Via del Fosso del Cavaliere 100, 00133 Roma, Italy \\
$^{3}$INAF-Osservatorio di astrofisica e scienza dello spazio di Bologna, Via Piero Gobetti 93/3, 40129 Bologna, Italy \\
$^{4}$Instituto Nacional de Astrof\'isica, Optica y Electr\'onica, Apartado Postal 51-216, 72000 Puebla, Mexico \\
$^{5}$Instituto de Radioastronom\'ia y Astrof\'isica (IRyA-UNAM), 3-72 (Xangari), 8701, Morelia, Mexico \\
$^{6}$Instituto de Astrof\'{i}sica de Andaluc\'{i}a, CSIC, Glorieta de
la Astronom\'{i}a, s/n, 18008 Granada, Spain\\
$^{7}$Argelander Institute for Astronomy, University of Bonn, Auf dem H\"{u}gel 71, D-53121 Bonn, Germany \\
$^{8}$International Max Planck Research School of Astronomy and Astrophysics at the Universities of Bonn and Cologne}

\begin{document}

\date{Draft:\today}

\pagerange{\pageref{firstpage}--\pageref{lastpage}} \pubyear{2019}

\maketitle

\label{firstpage}


\begin{abstract}
Mrk\,1498 is part of a sample of galaxies with extended emission line regions (extended outwards up to a distance of $\sim$7 kpc) suggested to be photo-ionized by an AGN that has faded away or that is still active but heavily absorbed. Interestingly, the nucleus of Mrk\,1498 is at the center of two giant radio lobes with a projected linear size of 1.1 Mpc. Our multi-wavelength analysis reveals a complex nuclear structure, with a young radio source  
(Giga-hertz Peaked Spectrum) surrounded by a strong X-ray nuclear absorption, a mid-infrared spectrum that is dominated by the torus emission, plus a circum-nuclear extended emission in the [OIII] image (with radius of $\sim$ 1 kpc), most likely related to the ionization of the AGN, aligned with the small and large scale radio jet and extended also at X-rays. In addition a large-scale extended emission (up to $\sim$ 10 kpc) is only visible in [OIII]. These data show conclusive evidence of a heavily absorbed nucleus and has recently restarted its nuclear activity. To explain its complexity, we propose that Mrk\,1498 is the result of a merging event or secular processes, such as a minor interaction, that has triggered the nuclear activity and produced tidal streams. The large-scale extended emission that gives place to the actual morphology could either be explained by star formation or outflowing material from the AGN.
\end{abstract}



\section{Introduction}

Active galactic nuclei (AGN) emit over the whole electromagnetic spectrum, and the observed emission at each frequency gives information about different physical regions. Thus the study of AGN gathered at different wavebands can be very useful in investigating the innermost parts of the galaxies, as well as the relation among them. 

The overall picture we have of AGN today is a supermassive black hole (SMBH) fed by an accretion disk which peaks at UV frequencies, that is thought to be surrounded by a discrete continuum of clouds distribution. Very close to the SMBH, located at about 0.01-1 pc, the clouds are moving at very high velocities ($v \sim $3000$ km/s$) composing the broad line region (BLR), responsible for the broad components in the optical spectra. A dusty structure (often simplified as a torus) of size around 1-10 pc encloses the BLR. This structure attenuates the optical/UV emission and re-emits it in the infrared. Therefore, mid-infrared frequencies are ideal to reveal the properties of the dust in the nuclear region related to the torus, as well as X-rays where the column density measured in the spectrum is thought to be related with absorption by the torus \cite[e.g.,][]{omaira2015}. X-rays are usually able to penetrate through the dusty torus, allowing to investigate coronal emission at few gravitational radii from the SMBH.
At kpc scales the narrow line region (NLR) is present, with clouds moving at $v \sim $500$ km/s$, the region responsible for the narrow lines in the optical spectra.
About 10\% of AGN also show strong relativistic jets that imprint their mark at radio frequencies \citep{kellermann1989}, although a continous distribution in radio powers suggests that AGN are radio emitters at any level \citep{panessa2019}.

One key question in understanding AGN is their evolution. The same galaxy can undergo through different processes where it accretes matter onto the SMBH to a status in which the accretion vanishes (i.e., a faded AGN), as well as to reactivate its nuclear activity and give place to a new born young AGN. The timescales involved in such processes can range between 10$^{4-8}$ years \citep{reynolds1997}. Given that the transitions occur in these timescales, targeting such systems becomes difficult because only a fraction of AGN might undergo intermediate phases.

Some galaxies show a radio morphology that exhibit linear spatial scales larger than 0.7 Mpc, they are classified as giant radio galaxies \citep[GRG,][]{ishwara1999}. These galaxies represent perfect laboratories to study the lifetime of nuclear activity. Because of the long time it takes the matter to travel all through the jets, starting from the nucleus to arrive to the lobes, this emission must be very old. Indeed, these timescales can be of the order of 10$^8$yr, thus different phases of nuclear activity might be observed in the same galaxy. The usual way to find restarting activity is through the radio morphology, where different phases of jet activity can be observed in the same image at different spatial scales \citep[e.g.,][]{schoenmakers2000}. However, when the restarted nuclear activity is in its earliest phase and is not still imprinted in the radio band, the detection of such young nuclei becomes a difficulty. 

In the present work we focus on the nucleus of Mrk\,1498 (also named WN\,1626+5153, 1626+518, Swift J1628.1+5145, RA=16h28m04.0s,  DEC=+51d46m31s, z=0.0547, \citealt{degrijp1992}). This is a FR II radio source showing a classical double-lobed structure and an unresolved nucleus, which was discovered in the Westerbork Northern Sky Survey (WENSS) at 325 MHz, while in the optical it shows a broad H$\alpha$ component (FWHM $\sim$ 6500 $km/s$), and therefore is classified as a Seyfert 1.9 \citep{rottgering1996,winter2010}, and is hosted in an elliptical galaxy.

Mrk\,1498 is included in the sample of \cite{bassani2016}, 
who performed a radio/gamma-ray study of the INTEGRAL-\emph{Swift} AGN population. Among the sample, 16 galaxies (25\%) are classified as GRG. We are developing a study of individual sources within these GRGs that show peculiar characteristics \citep[see][for PKS\,2331-240]{lore2017a, lore2018}. Interestingly, the radio lobes in Mrk\,1498 have a projected separation of $\sim$ 1.1 Mpc \citep{rottgering1996,schoenmakers2001}, while the orientations of the emission-line structures observed by the Hubble Space Telescope (\emph{HST}) and the large scale radio lobes observed by the Very Large Array (\emph{VLA}) differ by about 70 degrees \citep{keel2015}. This is unusual because these structures should be aligned in the nuclear region if AGN photoionization is responsible for the observed emission, where the jet is on the axis of symmetry of the torus \citep[e.g.,][]{Balmaverde2014,gomezguijarro2017}.

\cite{keel2012} selected this source on the basis of a sample of galaxies with AGN-ionized regions at projected radii larger than 10 kpc, based on the locations of the galaxies in standard diagnostic diagrams \citep[BPT,][]{baldwin1981}, strength of [Ne V] and He II emission lines, electron temperatures smaller than 20000 K, and that did not show strong signs of AGN obscuration based on far-infrared measurements ($L_{ion}/L_{FIR}$), i.e., consistent with photoionization but not with shock heating. The authors suggested that some of the extended emissions in their sample, composed by eight targets, could be interpreted as clouds ionized by an AGN which have faded over the differential light travel time between our views of the clouds and nuclei.

The Hubble Space Telescope (\emph{HST}) data of Mrk\,1498 shows extended emission in the [OIII] emission line image, revealing multiple clumps of ionised gas at radii 0.6-1.8 kpc kinematically dominated by rotation \citep{keel2017}, plus irregular and asymmetric dust lanes extending outward to 7.5 kpc. 
In the particular case of Mrk\,1498, the low ratio of $L_{ion}/L_{FIR}$=0.6 \citep{keel2012} is in contrast with the high value of mid-infrared luminosity, leading \cite{keel2015,keel2017} to infer that the extended emissions could be interpreted as clouds ionized by a faded AGN, but also left open the possibility of ionization cones within an obscured AGN.  In fact, at X-rays, data from \emph{Suzaku} has been used to report Mrk\,1498 as an obscured AGN \citep{eguchi2009,kawamuro2016}.

In the present work we use mutiwavelength data at radio, mid-infrared, optical, ultraviolet, and X-rays to disentangle whether a faded or an obscured AGN is present in the nucleus of Mrk\,1498, as well as to study the extended emission and the connection between the different structures observed at different frequencies. The paper is organized as follows: In Section 2 we present the data and its reduction for the different wavebands, whose analysis and results are reported in Section 3. This is followed by a discussion about the nuclear characteristics of the galaxy in Section 4. Finally, a summary of our results is presented in Section 5.

We adopt the latest cosmological parameters from the \emph{Planck} mission (\citealt{2018arXiv180706209P}), i.e. assuming the base-$\Lambda\rm{CDM}$ cosmology: $H_{0}=67.4$ km/s/Mpc, $\Omega_{\rm{m}}=0.315$, and $\Omega_{\Lambda}=0.685$.


\section{Data and its reduction}

In this section we present the data used in the analysis. We have compiled archival data at radio (VLA, VLBA, and Effelsberg), mid-infrared (\emph{Spitzer}), optical (San Pedro M\'artir Telescope, and Hubble Space Telescope), and X-ray (\emph{Chandra}, \emph{XMM--Newton}, \emph{Swift}, and \emph{NuSTAR}) frequencies.

\subsection{Radio}

We considered the NRAO VLA Sky Survey (NVSS) image at 1.4 GHz \citep{condon1998}, with an angular resolution of 45arcsec$\times$45arcsec: this was used by \cite{bassani2016} for the GRG classification of the parent sample. In addition to this, in 2017 the TIFR GMRT Sky Survey at 150 MHz was published (TGSS, \citealt{Intema2017}), providing a map of Mrk\,1498 at low frequency with an angular resolution of 25 arcsec $\times$ 25 arcsec. Archival data from the Very Large Baseline Array (VLBA, NRAO) are available at 4.8 GHz from February 2009, with the entire array (10 antennas). Moreover, we searched the Very Large Array (VLA, NRAO) archive, looking for imaging data at a better resolution than NVSS, to probe an intermediate scale between the VLBA and the NVSS one. We found data at 5 GHz from November 2015, with an angular resolution of $\sim$20$\times$10 arcsec. Finally, we considered the single-dish Effelsberg-100m data at 4.8, 8.4, and 10.5 GHz from our GRG observing campaign \citep{bruni2019}. These were used to reconstruct the core radio spectrum, given the low angular resolution (see Sect. \ref{res:radio}). 


\begin{figure*}
\includegraphics[width=20cm,trim={2cm 3cm 0 2cm},clip]{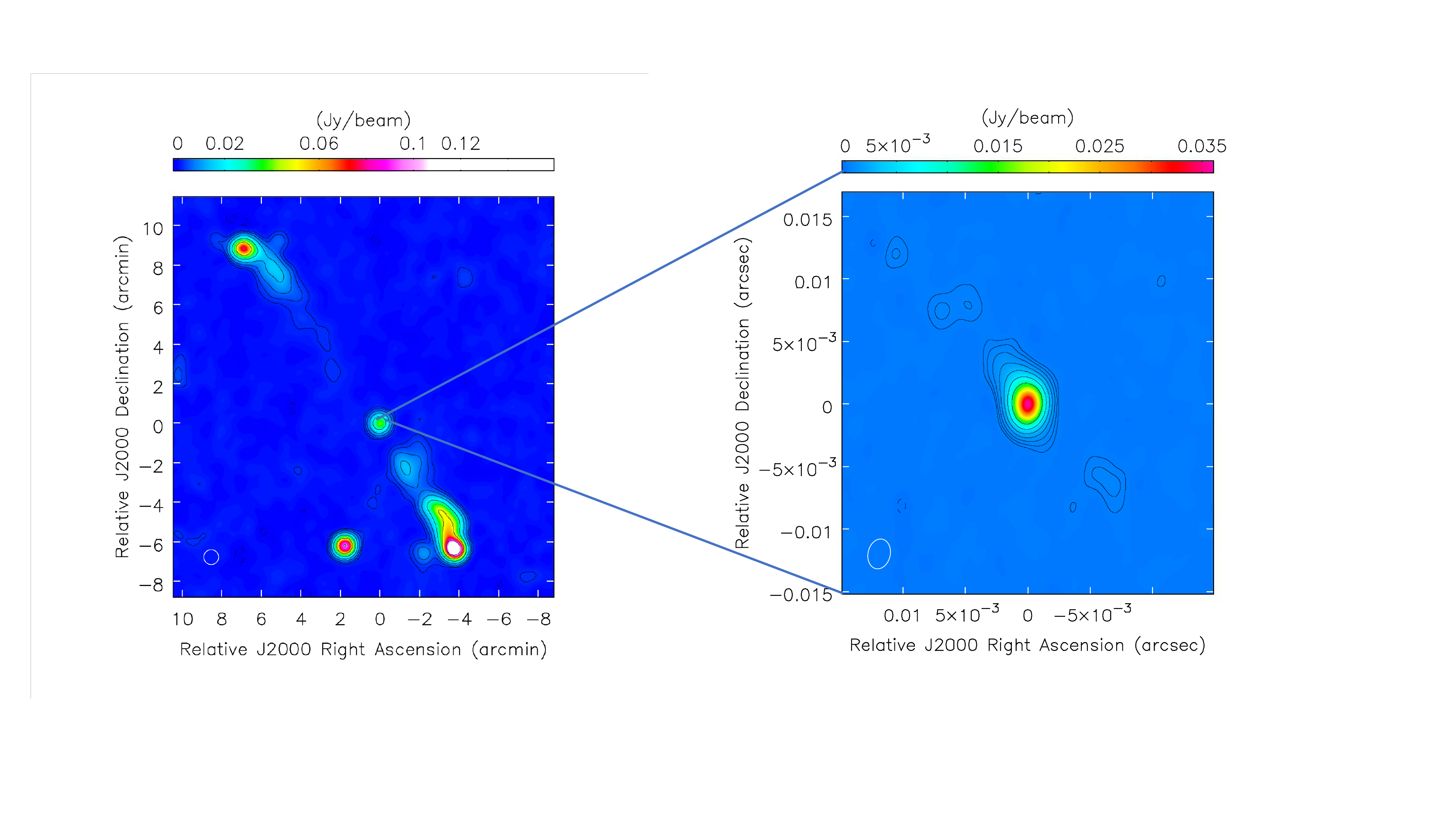}
\caption{(Left) NVSS image of Mrk1498, at 1.4 GHz. Contours are 3$\sigma$ multiples (-1, 1, 2, 4, 8, 16, 32, 64). The RMS is 0.5 mJy/beam, while the beam (45$\times$45 arcsec) is plotted in the lower-left corner, and (right) zooming image of the nucleus with VLBA at 4.8 GHz from archival data. Contours are 3$\sigma$ multiples (-1, 1. 2, 4, 8, 16, 32, 64). The RMS is 0.06 mJy/beam, while the beam (2.41$\times$1.78 mas) is plotted in the lower-left corner. Not at scale.}
\label{NVSS}
\end{figure*}


\subsection{Mid-infrared}

\emph{Spitzer} \citep{houck2004} observed Mrk\,1498 with the IRS instrument in the high resolution (HR) mode on 2009 March 5th within an observing program to obtain infrared Observations of a Complete Unbiased Sample of Bright Seyfert Galaxies (program ID 50253). We used the spectra from the Combined Atlas of Sources with \emph{Spitzer} IRS Spectra (CASSIS\footnote{http://cassis.sirtf.com}), which provide fully calibrated 1D \emph{Spitzer} spectra. No further calibrations were required. 


\subsection{Optical}

Optical spectra were obtained with the 2.1m telescope of the Observatorio Astron\'omico Nacional at San Pedro M\'artir (OAN-SMP), Baja California, M\'exico, on 2018 August 06th (start at 03:48 UT). Clear sky conditions were present, with a seeing of 2.5 arcsec. The Boller \& Chivens spectrograph was tuned to the 3700--8300 \AA\ range (grating 300 l/mm) with a spectral dispersion of 2.5 \AA/pix, corresponding to a full width at half maximum (FWHM) of 8.5 \AA\ derived from different emission lines of the arc-lamp spectrum. A 2.5 arcsec slit was used and oriented in the direction of the right ascension.
The spectral calibration was done by observing the spectrophotometric standard star Feige 110. 

The data reduction was carried out with the IRAF\footnote{IRAF is distributed by the National Optical Astronomy Observatories operated by the Association of Universities for Research in Astronomy, Inc. under cooperative agreement with the National Science Foundation.} software following standard procedures. The spectra were bias-subtracted and corrected with dome flat-field frames. Cosmic rays were removed interactively from all of the images. Arc-lamp (CuHeNeAr) exposures were used for the wavelength calibration. A spline function was fitted to determine the dispersion function (wavelength-to-pixel correspondence). Sky emission lines located at known wavelengths were removed during the wavelength calibration. The spectrum extraction aperture was 2.5$\times$50 arcsec.

Archival data from the Hubble Space Telescope (\emph{HST}) were available, observed on the 18th of May 2012 with the Advanced
Camera for Surveys (ACS) in narrow bands to isolate the [OIII]$\lambda$5007 (the FR551N filter) and H$\alpha \lambda$6563 (the FR716N filter) emission lines, and with the Wide Field Camera 3 (WFC3) to observe the stellar continuum in the F612M and F763M filters. These data have already been published in \cite{keel2015}, and we refer the reader to their work for details on the observations. We subtracted the continuum by following the levels of 0.10 for [OIII] and 0.15 for H$\alpha$ as in \cite{keel2015}.

\subsubsection{\label{optmon}Ultraviolet}

Taking advantage of the optical monitor onboard \emph{XMM--Newton} \citep[OM,][]{mason2001} and the Ultraviolet and Optical Telescope \emph{Swift} \citep[UVOT,][]{roming2005}, we performed optical and UV photometry that we used for the variability analysis. These data were taken simultaneously to those at X-rays in the six filters (UVW1, UVW2, UVM2, U, B, and V) on the
2007 June 23th with \emph{XMM--Newton} and on 2015 January 29th, February 9th, and May 21st with \emph{Swift}. 

We used the OM observation FITS source lists (OBSMLI\footnote{ftp://xmm2.esac.esa.int/pub/odf/data/docs/XMM-SOC-GEN-ICD-0024.}) to obtain the photometry with \emph{XMM--Newton}, and the {\sc uvotsource} task for \emph{Swift}, using a circular aperture
radius of 5 arcsec centred on the coordinates of the source.
The background region was selected as a free of sources circular
region of 20 arcsec close to the nucleus.


\subsection{X-rays}

The nucleus of Mrk\,1498 was observed by \emph{Chandra} on 2015 May 15th (ObsID. 17085)  with  the  ACIS  instrument \citep{garmire2003}.  The  data  reduction  and  analysis  were carried out in a systematic, uniform way using CXC \emph{Chandra} Interactive  Analysis  of  Observations  (CIAO),  version  4.9. Level 2  event data  were  extracted by using  the task {\sc acis-process-events}.  We  first  cleaned  the  data  from  background flares due to low-energy photons that interact with the detector. To clean them we used the task \texttt{lc\_clean.sl}\footnote{http://cxc.harvard.edu/ciao/ahelp/lc\_clean.html}, which removes periods of anomalously low (or high) count rates from light curves from source-free background regions of the CCD. 

Nuclear spectra were extracted from a circular region centred on the  positions  given  by  NED\footnote{https://ned.ipac.caltech.edu/}, using a circular region of 2 arcsec for the nucleus and of 7 arcsec for the background. We  used  the {\sc dmextract}
task  to  extract  the  spectra  of the  source  and  the  background  regions.  The response matrix file (RMF) and ancillary reference file (ARF) were generated for each source region using the {\sc mkacisrmf} and {\sc mkwarf} tasks, respectively.

The source was observed by \emph{XMM--Newton} on 2007 June 23th (ObsID. 0500850501). We used the data from the EPIC pn camera \citep{struder2001}. 
The data were reduced using  the  Science  Analysis  Software  (SAS\footnote{https://www.cosmos.esa.int/web/xmm-newton/sas}),  version 17.0.0.
This observation was affected by high background flares, thus we performed an optimization of the signal-to-noise ratio (S/N) in the 0.3-10 keV range in order to recover the maximum possible information of the observation \citep[see][]{piconcelli2004}. The background was extracted from a circular region with a radius of 50 arcsec, while the source extraction radius was allowed to be in the range 20--40 arcsec. We obtained the highest S/N of 50 with a source radius of 14 arcsec, recovering a net exposure time of 7.5 ks.

Mrk\,1498 was also observed by \emph{Swift} on 2015 January 29th (ObsID. 00035271001), February 9th (ObsID. 00035271002), and May 21st (ObsID. 00081194001). The data reduction of the Swift X-ray Telescope (XRT, \citealt{burrows2005}) in the Photon Counting mode was performed by following standard routines described by the UK Swift Science Data Centre (UKSSDC) and using the software in HEASoft version 6.23. Calibrated event files were produced using the routine {\sc xrtpipeline}, accounting for bad pixels and effects of vignetting, and exposure maps were also created. Source and background spectra were extracted from circular regions with 30 arcsec and 1 arcmin radius. The {\sc xrtmkarf} task was used to create the corresponding ancillary response files. The response matrix files were obtained from the HEASARC CALibration DataBase. 

The source was also observed by \emph{NuSTAR} on 2015 May 11th (ObsID. 60160640002), with a net exposure time of 24 ks. The \emph{NuSTAR} data were reduced using the standard pipeline {\sc nupipeline} in the NuSTAR Data  Analysis  Software  ({\sc nustardas}, v1.8.0), using calibration files from NuSTAR caldb v20180312. We extracted the spectra using the standard tool {\sc nuproducts} for the
source data from circular regions with a radius of 75 arcsec, and the background from a blank area close to the source. The spectra were binned to have a S/N ratio greater than five in each spectral channel, and not oversampling the instrumental resolution by a factor greater than 2.5. The spectra from FPMA and FPMB were analysed jointly, but not combined.
In all cases, before  background subtraction,  the  spectra  were binned to have a minimum of 20 counts per spectral bin, to be able to  use  the $\chi^2$-statistics.  This was made  with the
{\sc grppha} task included in ftools.

Finally, we included the average 105-month \emph{Swift}/BAT spectra
from the most recent hard X-ray survey \citep{oh2018}.

For imaging, we used \emph{Chandra} data because it has the highest spacial resolution. We extracted \emph{Chandra} data in the soft (0.5-2.0 keV) and hard (2.0-10.0 keV) energy bands. The {\sc csmooth} task included in CIAO was used to adaptively smooth the images, using a fast Fourier transform algorithm and a minimum and
maximum significance level of the signal-to-noise of 3 and 4, respectively.


\section{Analysis and results}

\subsection{\label{res:radio}Radio}

The NVSS image at 1.4 GHz is shown in the left panel of Figure \ref{NVSS}. We measured a linear size of 1.2 Mpc (1125 arcsec from lobe to lobe, at a spatial scale of 1.087 kpc/arcsec) so it falls in the GRG category. Indeed, it is included in the recently-published sample of GRG selected in the soft gamma-ray band by \cite{bassani2016}, and classified as an FRII. The symmetric structure has a position angle of about 45 degrees, with the southern jet less extended than the Northern one, which could be due to a projection effect.

In the right panel of Figure \ref{NVSS} we present the VLBA image at 4.8 GHz, reaching an RMS of 0.06 mJy/beam. The core is clearly detected, with a peak flux density of 36 mJy/beam, and it shows an elongation towards NE/SW, at a position angle of about 45 degrees, consistent with the one detected at kpc scale from NVSS. With a restored beam of 2.41$\times$1.78  milliarcsecond (mas), the VLBA observation shows the central pc-scale structure of the AGN, thus we are detecting the most recent emission due to the accelerated plasma in the inner part of the jet. From a 2D Gaussian fitting of the central component visible in the image, we could estimate a deconvolved size of 1.3 mas for the major axis, while the minor one remains unresolved. At the redshift of the source, this translates into a projected linear size of 1.4 pc.

\begin{figure}
\includegraphics[width=0.45\textwidth]{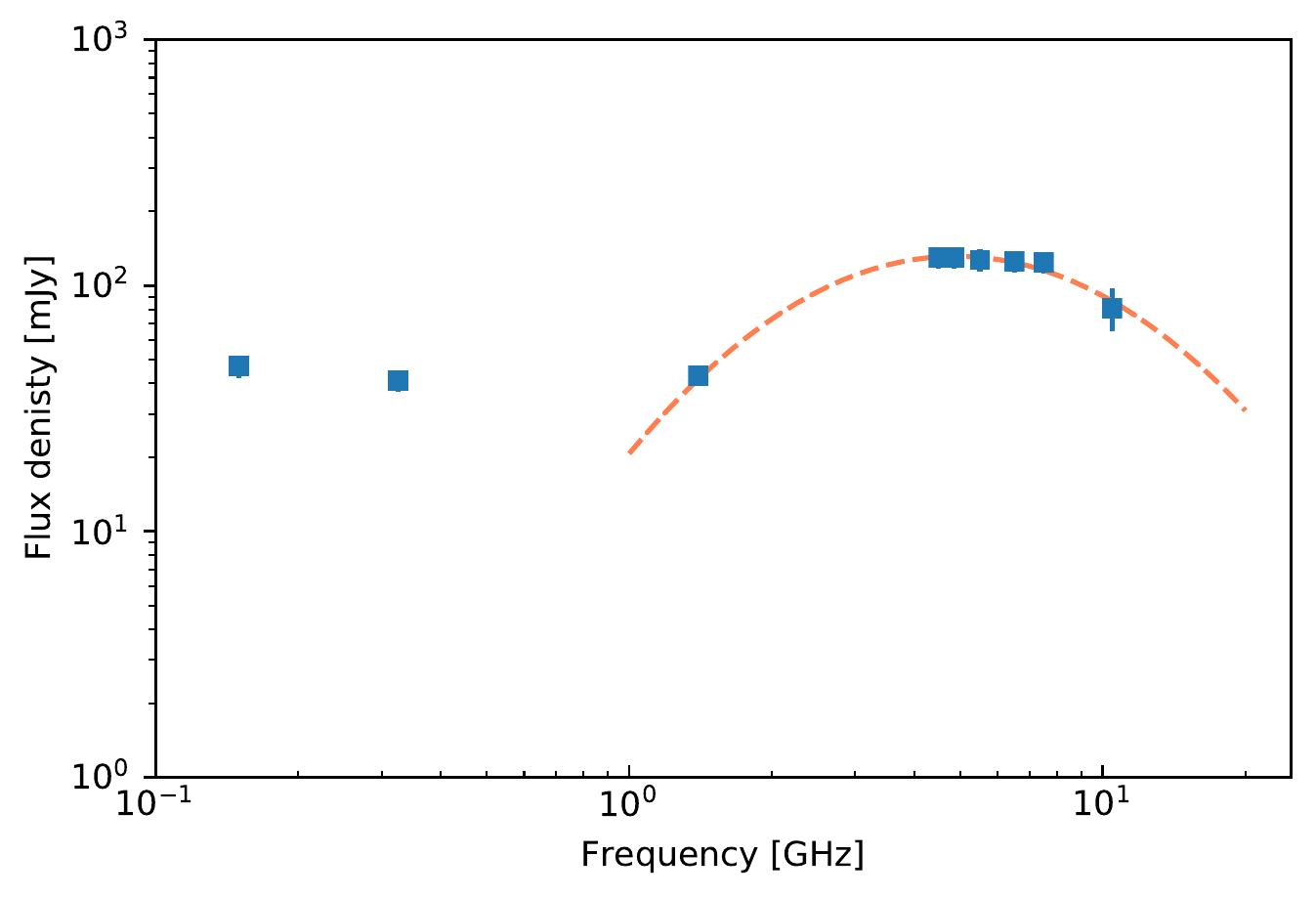}
\caption{Radio spectrum of the nucleus of Mrk 1498 from 150 MHz to 10 GHz, build with data from TGSS, WENSS, NVSS, Effelsberg-100m, and  EVLA \citep{bruni2019}. A log-parabola is fitted to show the spectrum curvature and peak in the GHz range.}
\label{Eff}
\end{figure}

Moreover, \cite{bruni2019} presented the nuclear radio spectrum using data from the TGSS, NVSS, Effelsberg-100m, and EVLA, that is shown in Figure \ref{Eff} (taken from \citealt{bruni2019}). A log-parabola model is fitted in the GHz-range, peaking at 4.9 GHz. The spectral shape of the fitting resembles those that characterize Giga-Hertz Peaked sources \citep{fanti1995}, which probably represent radio galaxies in the early stage of their life \citep{odea1998}.

\subsection{Mid-infrared}

In order to perform continuum spectral fit to the mid-infrared spectra we converted the \emph{Spitzer}/IRS spectrum into XSPEC format using the {\sc flx2xsp} task within HEASOFT. This tool reads a text file containing one or more spectra and errors and writes out a standard XSPEC pulse height amplitude (PHA) file and response file. These files are easily read by XSPEC to perform statistical tests for comparison between models. We have also used an additive table in XSPEC format for dust models using a clumpy \citep{nenkova2008} and a smooth \citep{fritz2006} distribution of clouds. A one-parameters table (in fits format) associated to all the spectral energy distributions (SED) is created using the {\sc flx2tab} task within HEASOFT. Headers are then changed to associate each SED to a set of parameters using a python routine. This model and data conversion in XSPEC format are widely used by Gonzalez-Martin et al. (submitted) and Esparza-Arredondo et al. (submitted). 

The parameters of the clumpy model are the viewing angle toward the torus, $i$, inner number of clouds, $\rm{N_{0}}$, the width of the torus, $\rm{\sigma}$, outer radius of the torus, $\rm{R_{out}}$ (compared to the inner radius, i.e,. $\rm{Y=R_{out}/R_{in}}$), the slope of the radial density distribution, $\rm{q}$, and the optical depth of the individual clouds at 5500 $\AA$, $\rm{\tau_{v}}$. The parameters of the smooth model are the viewing angle toward the toroidal structure, $i$, the opening angle, $\rm{\sigma}$, exponent of the logarithmic azimuthal density distribution, $\rm{\gamma}$, exponent of the logarithmic radial profile of the density distribution, $\rm{\beta}$, outer radius of the torus, $\rm{R_{max}}$, compared to the inner one (expressed as $\rm{Y = R_{out}/R_{min}}$), and the silicate dust distribution, $\rm{\tau_{9.7\mu m}}$. 

Contrary to most \emph{Spitzer}/IRS spectra of AGN (Gonzalez-Martin et al. submitted), either the smooth or the clumpy model is enough to reproduce the data; no additional circumnuclear emission from either dust heated by star-formation or stellar component are needed by the data (the addition of these two components were rejected by the F-test probability). This is in agreement with the lack of poly-aromatic hydrocarbon (PAH) features in the spectrum. We further notice that the fact that the mid-infrared luminosity is larger than the far-infrared (where star formation peaks) also discards star formation \citep{keel2017}. 

\begin{figure*}
\includegraphics[width=1.\columnwidth]{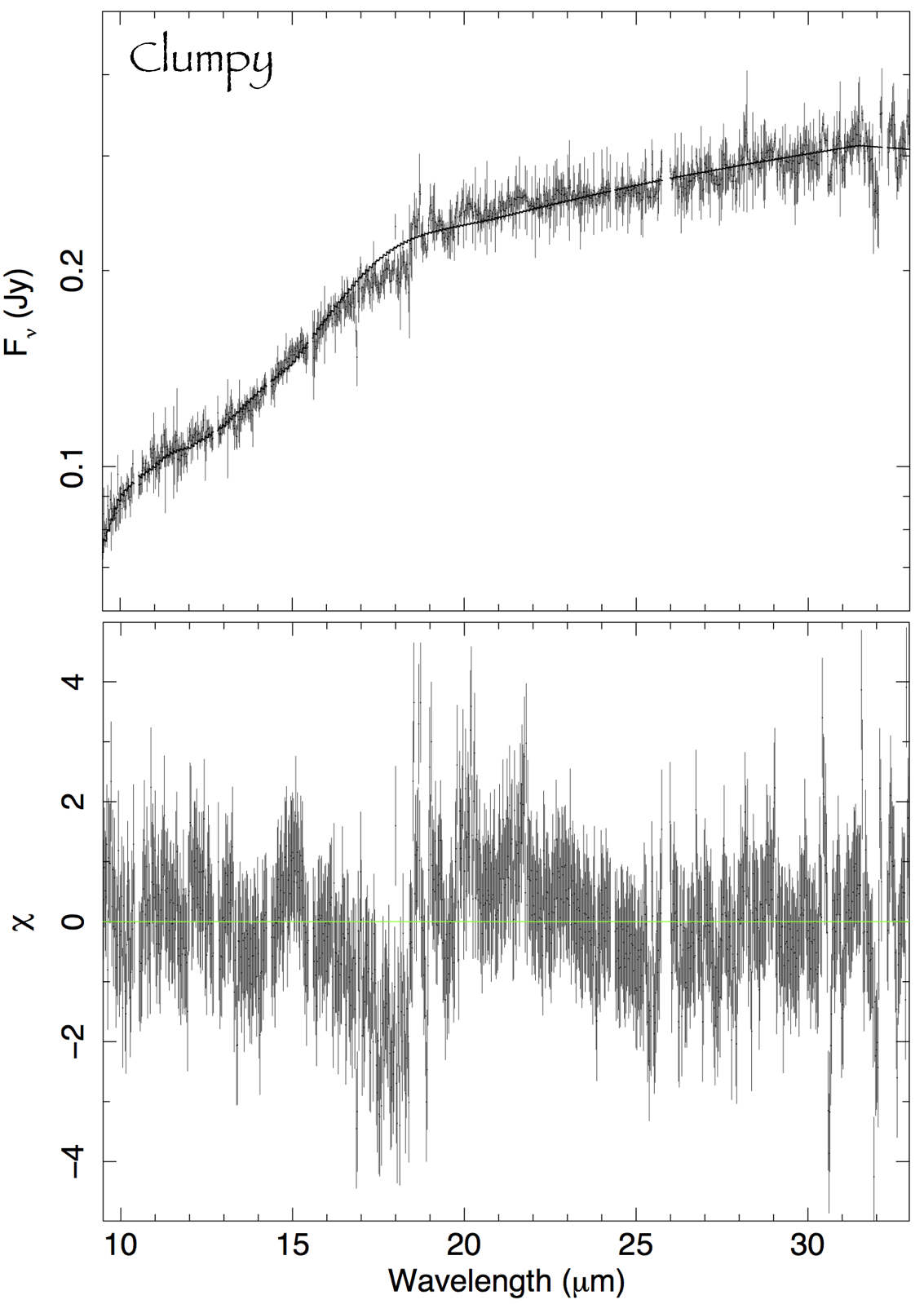}
\includegraphics[width=1.\columnwidth]{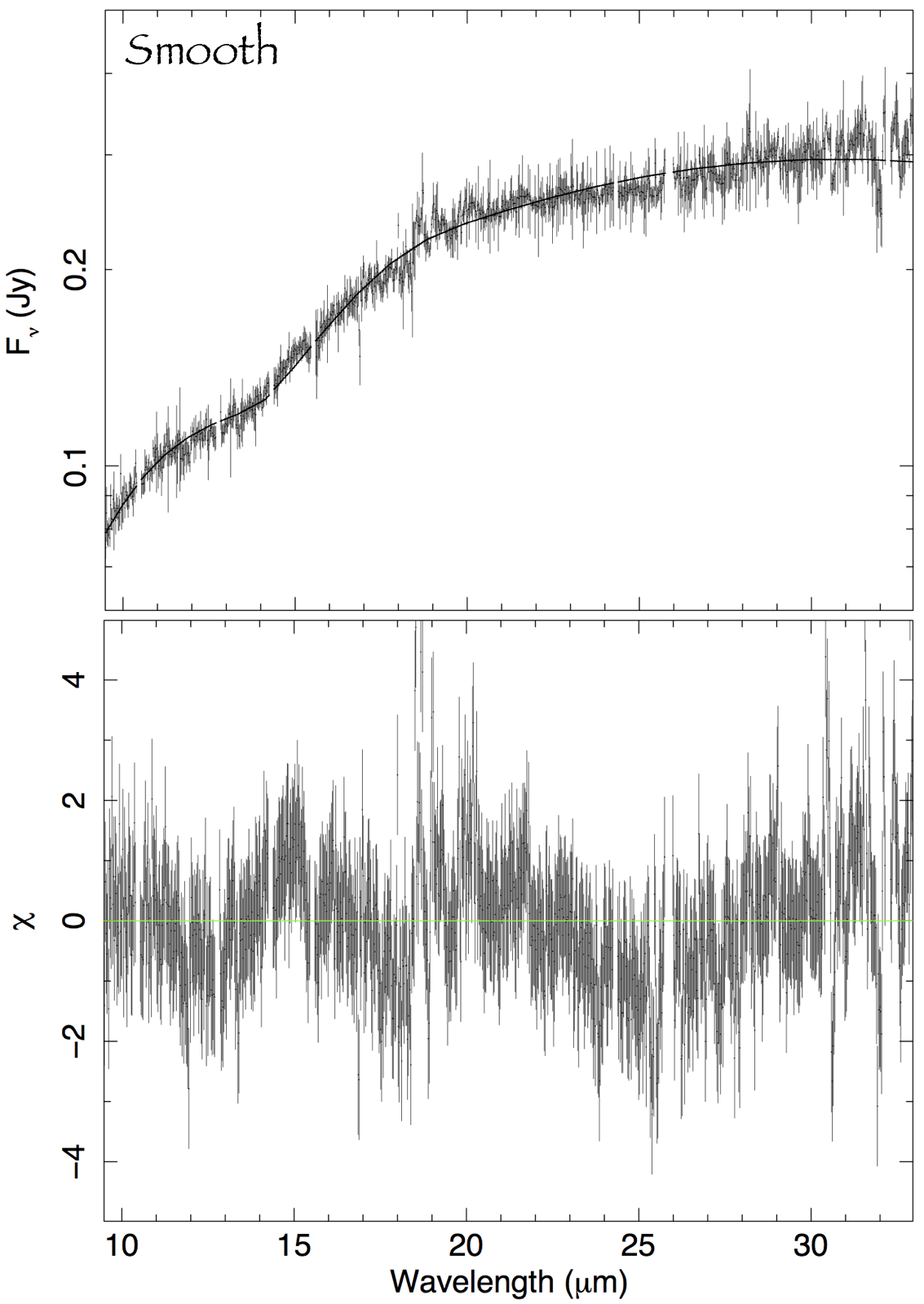}
\caption{ Spectral fit (top) and residuals (bottom) for the \emph{Spitzer} spectrum of Mrk\,1498. The left panel shows the best fit to Clumpy model by \citet{nenkova2008} and the right panel the best fit to the smooth model by \citet{fritz2006}. The spectrum is constituted by short high spectral resolution segments. The ensemble of these segments results in cosmetic issues \citep{houck2004}. Emission lines from Neon, Sulfur and Oxygen are ignored in the fitting procedure in order to better model the dust continuum emission.}
\label{fig:midinfraredfit}
\end{figure*}

The best fits and set of parameters are recorded in Figure\,\ref{fig:midinfraredfit} and Table\,\ref{tab:midinfraredfit}. Note that prominent emission lines from Neon, Sulfur and Oxygen are ignored in the fitting procedure in order to better model the dust continuum emission. Both clumpy and smooth models produce a reasonable fit with a $\rm{\chi^2/dof}$ of 1.12 and 1.20, where dof refers to the degrees of freedom. Both models produce a consistent picture in which the AGN is seen almost edge on according to the width of the torus and the viewing angle\footnote{Note that the viewing angle is defined from the pole of the torus in the Clumpy model while from the equator of the torus in the Smooth model.} since the angular width of the torus is small enough to produce an unobstructed view of the inner parts. 
The covering factor of the source, according to the Clumpy model is $\rm{f_{cov} = 0.34}$, while it is 0.40$\pm$0.02 in the smooth model.
The optical depths are $\rm{\tau =87.9_{-8.7}^{+11.8}}$ for the clumpy model, whereas only a lower limit can be estimated in the smooth model, with $\rm{\tau>9.8}$.

\begin{table}
\begin{center}
 \caption{\label{tab:midinfraredfit}Mid-infrared spectral fit results. Note that the viewing angles in both models are related as $\rm{i_{clumpy} = 90 - i_{smooth}}$ while half viewing angles of the torus use the same definition (i.e. measured from the equatorial plane of the torus). These are measured in degrees. The errors represent 1$\sigma$ of confidence levels.}
 \begin{tabular}{@{}cc}
  \hline
Clumpy & Smooth \\ \hline
$\rm{i = 70.0_{-1.3}^{+1.3}}$       &   $\rm{i = 32.9_{-1.3}^{+1.7}}$    \\     
$\rm{N_{0} = 7.6_{-0.5}^{+0.5}}$    & $\rm{\gamma > 5.4}$      \\
$\rm{ \sigma<15.3}$                 & $\rm{ \sigma  = 27.5_{-2.3}^{+2.4}}$      \\
$\rm{Y =30_{-4}^{+4}}$              & $\rm{Y =14.6_{-0.3}^{+0.2}}$      \\
$\rm{q =2.29_{-0.10}^{+0.05}}$      &  $\rm{\beta >-0.01}$      \\
$\rm{\tau =87.9_{-8.7}^{+11.8}}$    &  $\rm{\tau>9.8}$     \\
  \hline
 \end{tabular}
 \medskip
 \end{center}
\end{table}

\subsection{Optical}

\begin{figure*}
\includegraphics[width=\textwidth]{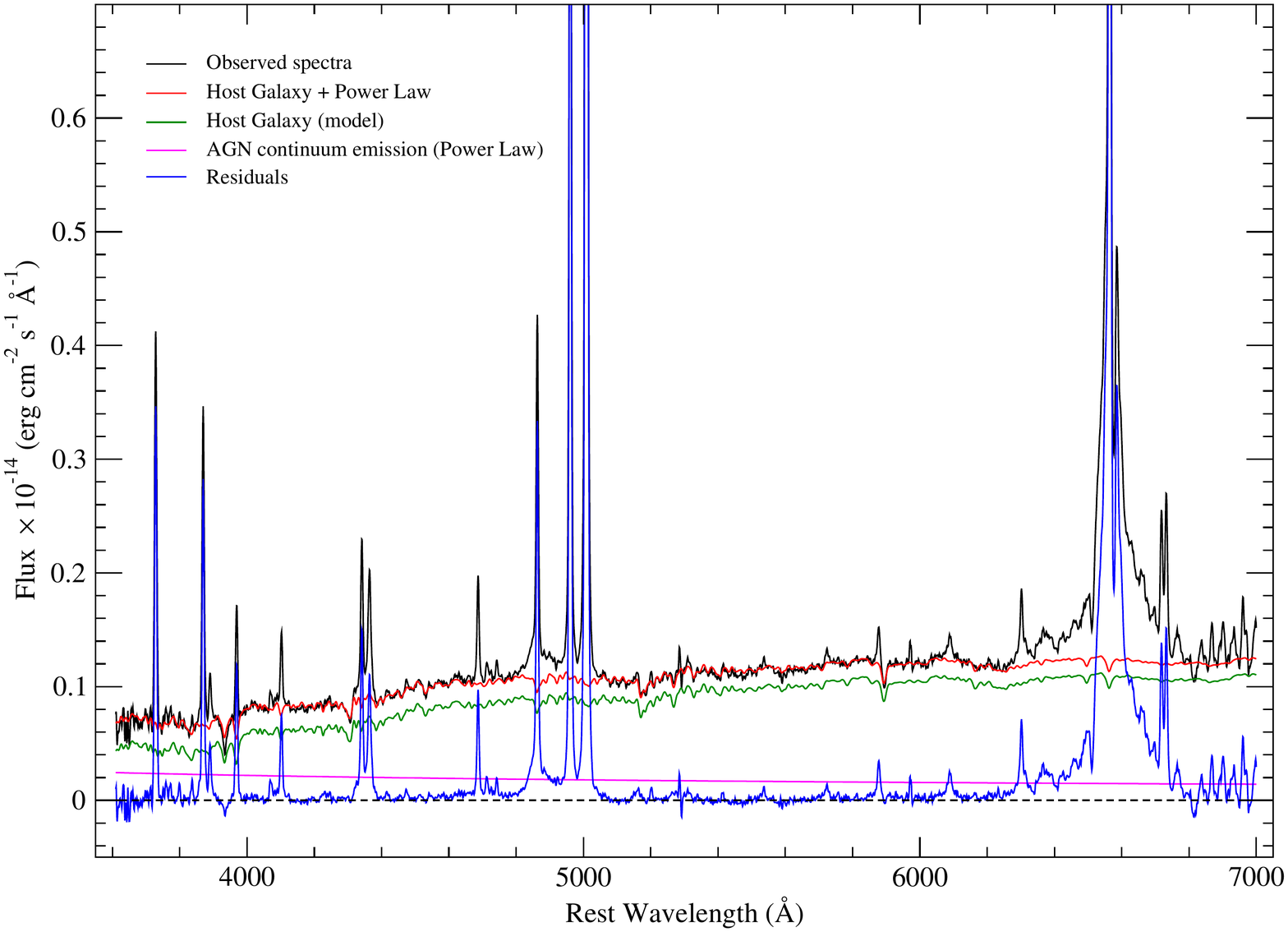}
\caption{ Decomposition of the optical spectrum of Mrk\,1498 taken with the SPM. The black corresponds to the original spectrum, the red to the best-fitted model, the green to the host galaxy contribution, the pink the continuum from the AGN emission, and the blue to the stellar population subtracted spectrum (i.e. observed -- best-fitted model). }
\label{spmstarlight}
\end{figure*}

We applied the STARLIGHT stellar population synthesis code \citep{cid2005} for the recovery of the shape of the stellar continuum to the SPM spectrum. We performed the spectral fitting using 150 single stellar populations from the evolutionary synthesis models of \cite{bruzual2003} and six different power-law slopes to simulate the AGN component to the optical continuum emission. The extinction law of \cite{cardelli1989} with Rv = 3.1 was adopted. The base components comprise 25 ages between 1 Myr and 18 Gyr, and six metallicities, from Z=0.005 $Z_{\odot}$ to 2.5 $Z_{\odot}$. The regions affected by atmospheric absorption and H$\alpha$+[NII] and H$\beta$+[OIII] emission line complexes were masked and excluded from the fit. The decomposed spectrum is presented in Figure \ref{spmstarlight}, where the subtracted spectrum is shown in blue.

\begin{figure}
\includegraphics[width=7cm]{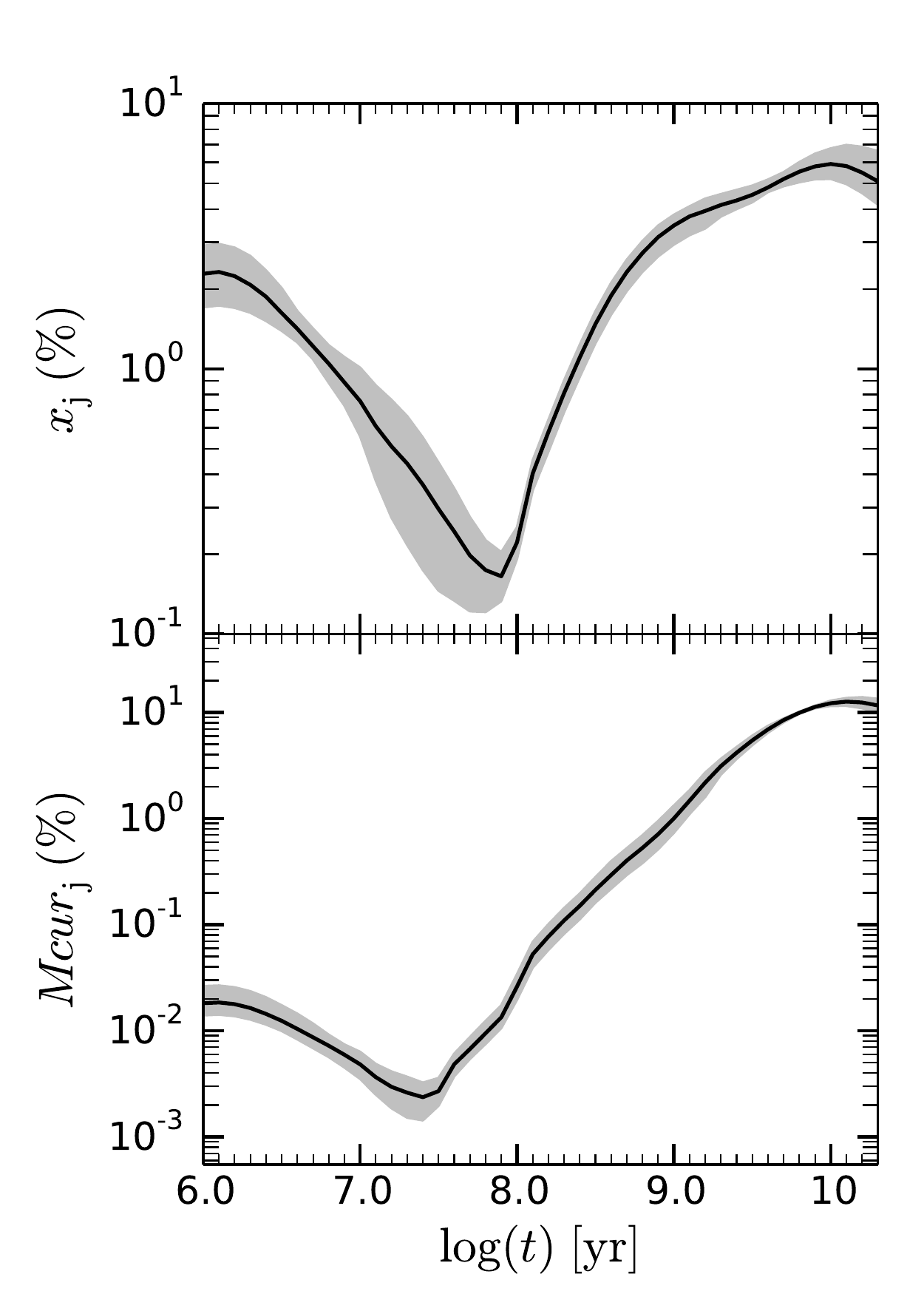}
\caption{SFH of Mrk\,1498 obtained from STARLIGHT. In the upper panel  the fractional percentage of stellar light as a function of its respective age is presented. In the lower panel, the fraction of the current mass and its respective age. The median trend (solid line) and its associated dispersion (16th-84th percentile, grey shaded region) are also presented.}
\label{sfh}
\end{figure}

We also used the output from STARLIGHT to gain insights on the star forming history (SFH). Figure \ref{sfh} shows the current light, $x_j$, and mass, $Mcur_j$, fraction as a function of the age of the j-esim component used in the stellar synthesis analysis. We have performed 100 realizations in order to derive the median trend (solid line) and its associated dispersion (16th-84th percentile, grey shaded region). The light and mass fraction indicates that the galaxy might have experienced a recent burst ($\sim10^6$ years old) of star formation, that correspond to the current 0.01-0.1\% of the galaxy’s stellar mass, M$_{\star}$. We note that the AGN non-thermal component might be ``diluting'' the stellar absorption lines, which hinders the accuracy of the spectral synthesis analysis and thus the values reported above should be treated with caution.

We estimated also the stellar velocity dispersion, $\sigma_{\star}$. The observed value is 
221$\pm$2 km s$^{-1}$, whereas the instrumental broadening is 183$\pm$7 km s$^{-1}$, thus we obtained $\sigma_{\star}$=124$\pm$11 km s$^{-1}$. Using this measurement and following the relation in \cite{gultekin2009}, we inferred a black hole mass  of  $\log(M_{\rm BH}^{\sigma_{\star}}/{\rm M}_{\odot})=7.2 \pm 0.2$.  

After the subtraction of the stellar contribution, each of
the emission lines H$\beta$, [OIII], [OI], H$\alpha$-[NII] and [SII] in
the spectrum were modelled with single or multiple Gaussian-profiles with
a Levenberg-Marquardt least-squares fitting routine (\texttt{MPFITEXPR},
implemented by \citealt{Markwardt2009}) within the Interactive Data
Analysis
environment\footnote{http://www.harrisgeospatial.com/SoftwareTechnology/IDL.aspx
(IDL)}. We have imposed that the intensity ratios between the [N
II]$\lambda$6548 and the [N II]$\lambda$6583 lines, and the [O I]
$\lambda$6363 and the [O I] $\lambda$6300 lines satisfied the 1:3 and
1:2.96 relations, respectively \citep{Osterbrock2006}. [OIII] features
follow the same relation as the [NII] lines. 

Following \citet{Cazzoli2018}, we estimated the accuracy of the fit as
the ratio between the standard deviation values estimated from a
wavelength range under the \textit{line} of interest $\varepsilon_{\rm
line}$ and that from a portion of the continuum free of both emission
and absorption lines, $\varepsilon_{\rm c}$, i.e. $\varepsilon^{\rm
fit}_{\rm line}$ = $\varepsilon_{\rm line} / \varepsilon_{\rm c}$.

We fit the integrated spectrum in two spectral regions
(i.e. H$\beta$, [OIII] and [OI], H$\alpha$-[NII], [SII]). We first model
H$\beta$ and [OIII] lines, in the blue part of the spectrum, as these
are less affected by sky lines and also not blended, contrary to the
H$\alpha$-[NII] complex.

\begin{figure}
\includegraphics[width=9.5cm]{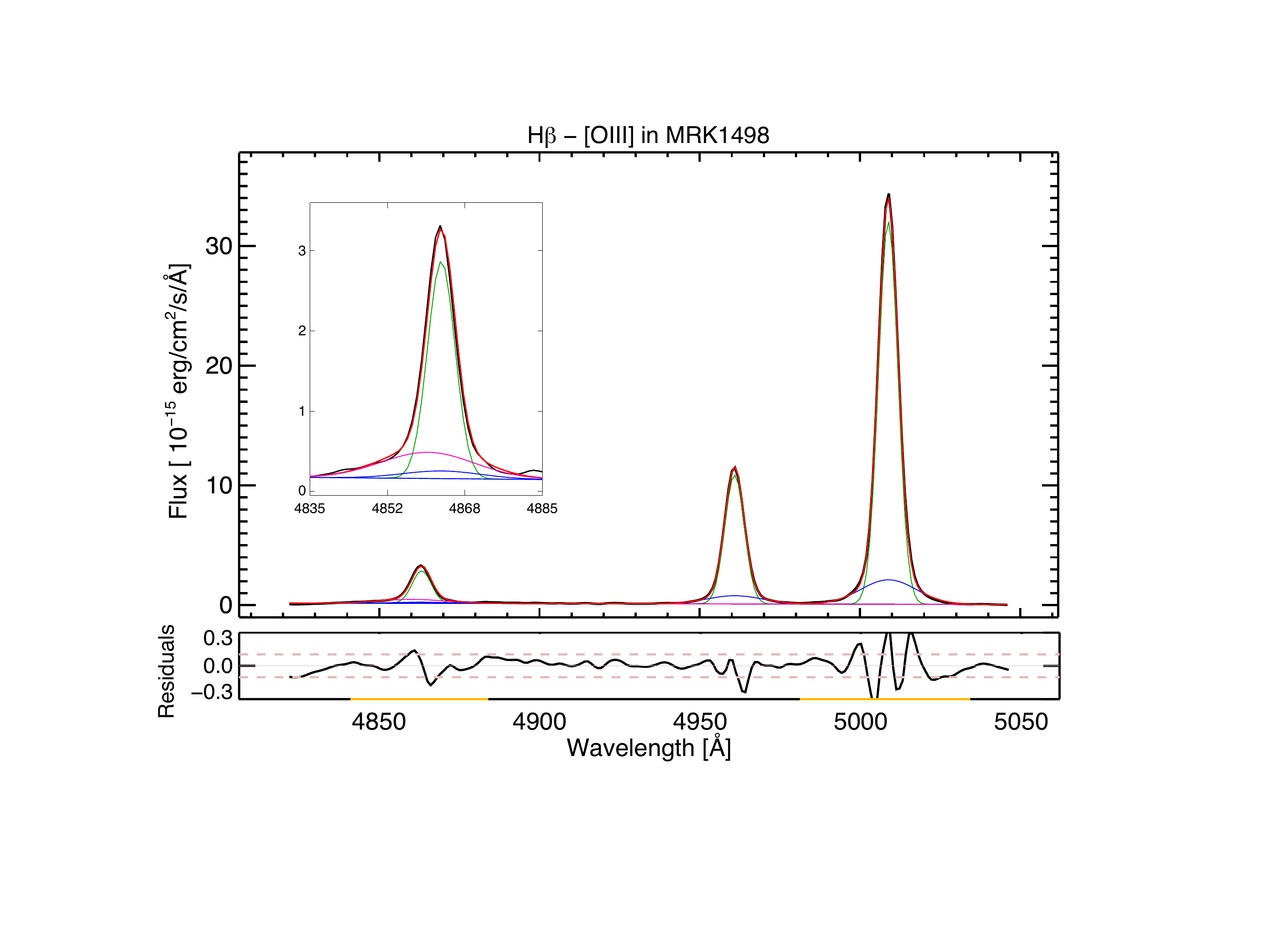}
\caption{Gaussian fits to emission lines profiles (after stellar
continuum subtraction) in the blue spectra. We Mrked with different
colours the Gaussian components required to model emission lines (same
colours Mrk same kinematic components). Specifically,  in green, blue
and pink the narrow, intermediate and broad components, respectively. In
light blue, the rest frame wavelengths.  The red curve univocally shows
the total contribution coming from the Gaussian fit.  Residuals from the
fit are in the small lower-panels in which pink lines indicate the
$\pm$\,3$\varepsilon$ (dashed) and the zero level (continuous). In
yellow, we Mrked the region used for the $\varepsilon$-test. The inset panel highlights the H$\beta$ region with a zoomed view.}
  \label{F_LF}
\end{figure}

After an accurate experimentation we find that the two Gaussians per
line led to a good fit of the [OIII] lines reducing significantly the
residuals and the values of $\varepsilon^{\rm fit}_{\rm H\beta}$ and
$\varepsilon^{\rm fit}_{\rm [OIII]}$ (from 4.5 and 15.3 to 3.5 and 3.9,
respectively) compared to a single Gaussian model. 
We classified these two components according to their line width as
narrow and intermediate. Then, we  explore the possibility of a broad
component in H$\beta$. By adding a broad component $\varepsilon^{\rm
fit}_{\rm H\beta}$  decreases (2.6), significantly improving the spectral fit. The results are shown in
Figure\,\ref{F_LF} and summarized in Table\,\ref{T_LF}.

We did not add any third component in [OIII] lines (which would be
difficult to explain physically) and we consider the line fitting of
these lines accurate to a 4$\varepsilon_{\rm c}$ level.  A deeper spectrum
or data at higher spectral resolution are needed to confirm the presence
of any additional components in [OIII]. Thus we can classify this source as a Seyfert 1.8 according to \citet{osterbrock1981}. 

For the lines in the red part of the spectrum, we first tried to fix the kinematic parameters (velocity shift and width) according to the
H$\beta$-[OIII] fits.
However, the results of this test were quite unsatisfactory, this might
be due to the combination of the severe H$\alpha$-[NII] blend and the contamination from sky lines that affect the lines profiles.
We also performed the fits in the red spectral region independently to that in the blue, also considering the [OI] lines as template for the H$\alpha$-[NII] blend. Although weak, these lines are not affected by the broad component in H$\alpha$, in contrast to [SII] that are contaminated by a broad red tail. However,  none of these tests were satisfactory, we therefore consider the fit in the red part of the spectrum unreliable and hence we will focus only on the results from the fit of the H$\beta$ and [OIII] lines. \\

\begin{table}
\caption{Results from the modeling of optical emission lines. The measurements are corrected by instrumental broadening.}
\begin{tabular}{l c c}
\hline
component & V & $\sigma$   \\
                   &  kms$^{-1}$ &    kms$^{-1}$    \\
    \hline
  Narrow         & 118.2\,$\pm$\,13 & 33.4\,$\pm$\,7   \\
  Intermediate & 124.4\,$\pm$\,9 & 485.3\,$\pm$\,14 \\
  Broad           & -61.7\,$\pm$\,33& 590.2\,$\pm$\,74 \\
\hline
\end{tabular}
\label{T_LF}
\end{table}


\begin{figure*}
\includegraphics[width=0.33\textwidth]{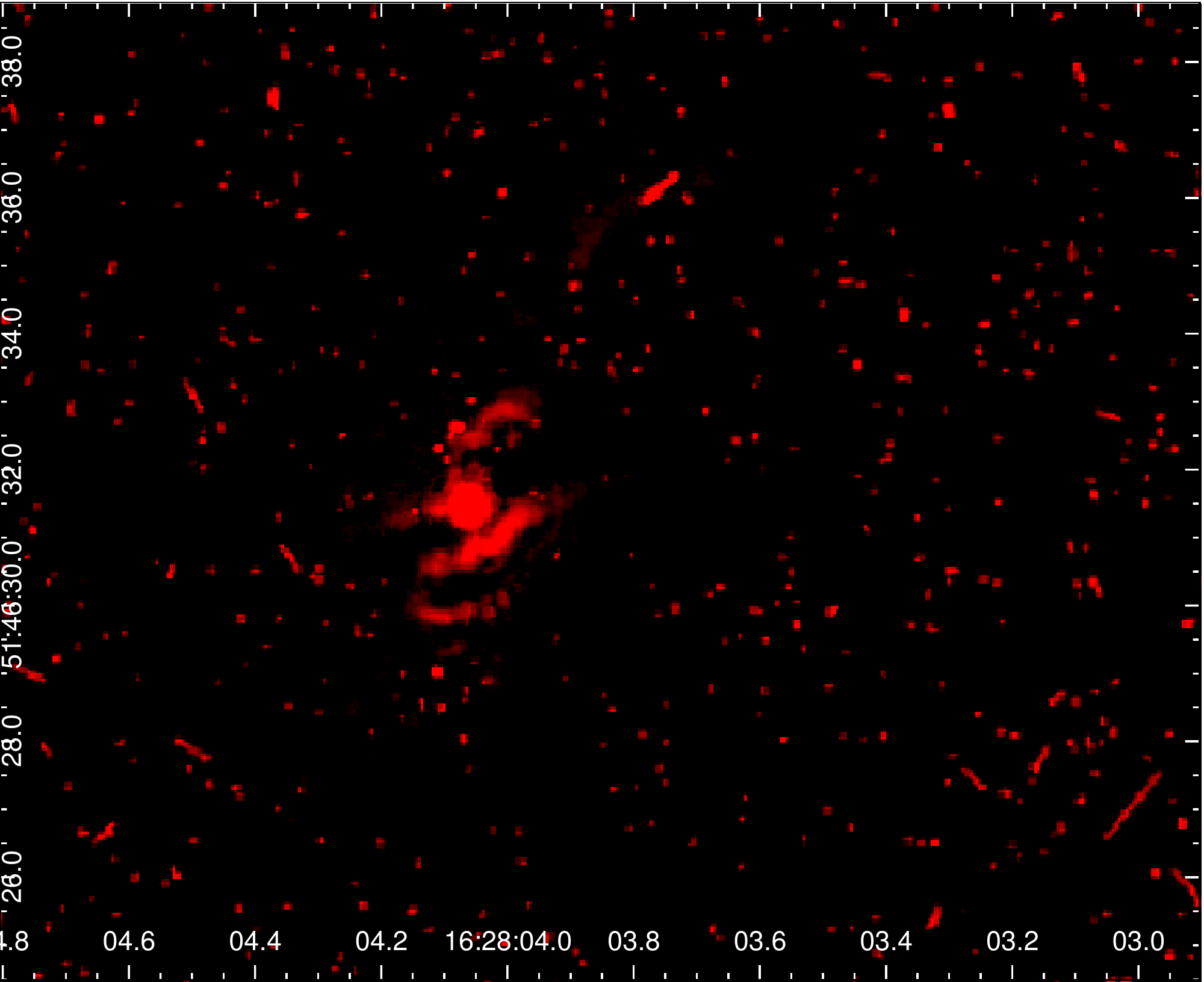}
\includegraphics[width=0.33\textwidth]{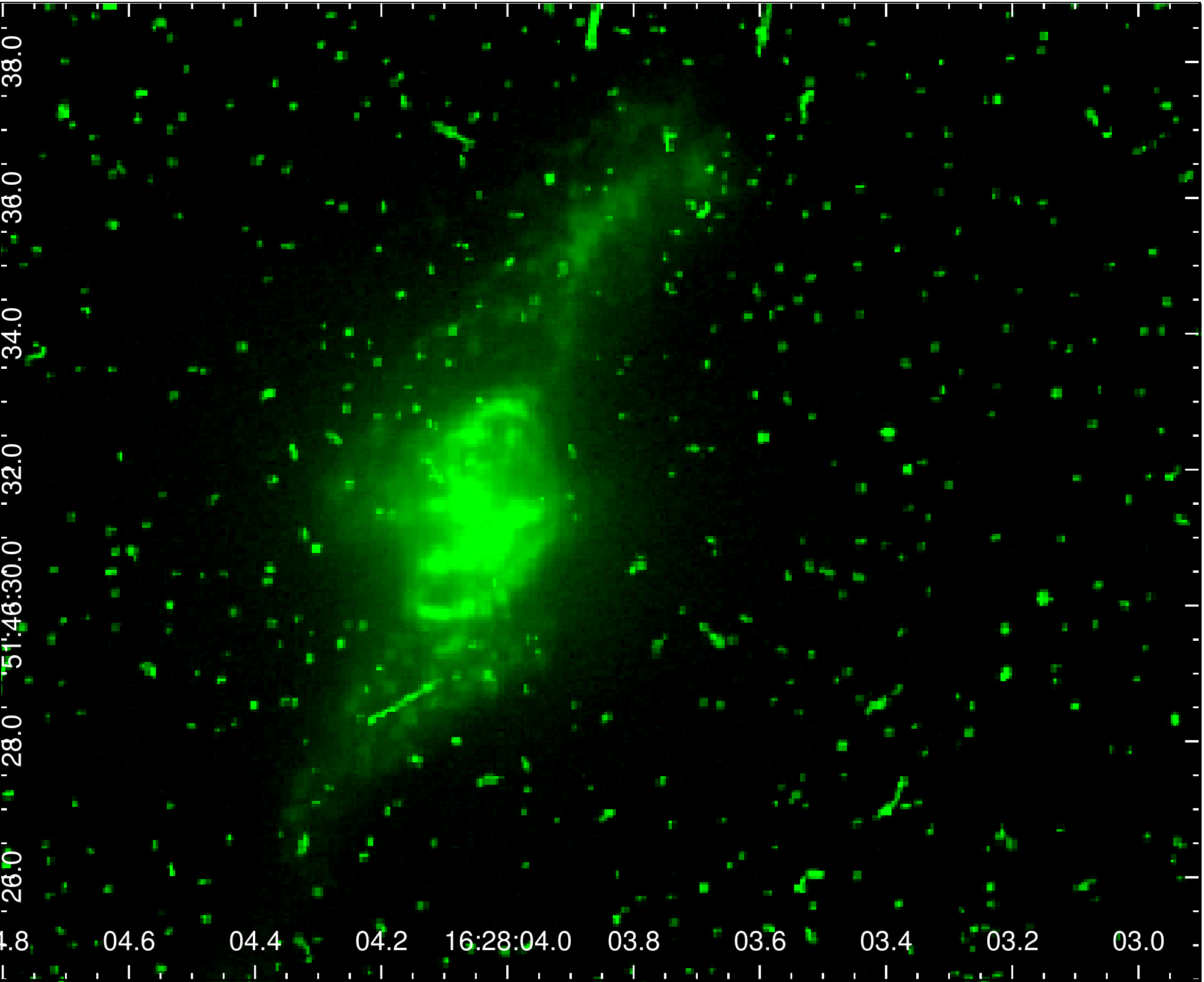}
\includegraphics[width=0.33\textwidth]{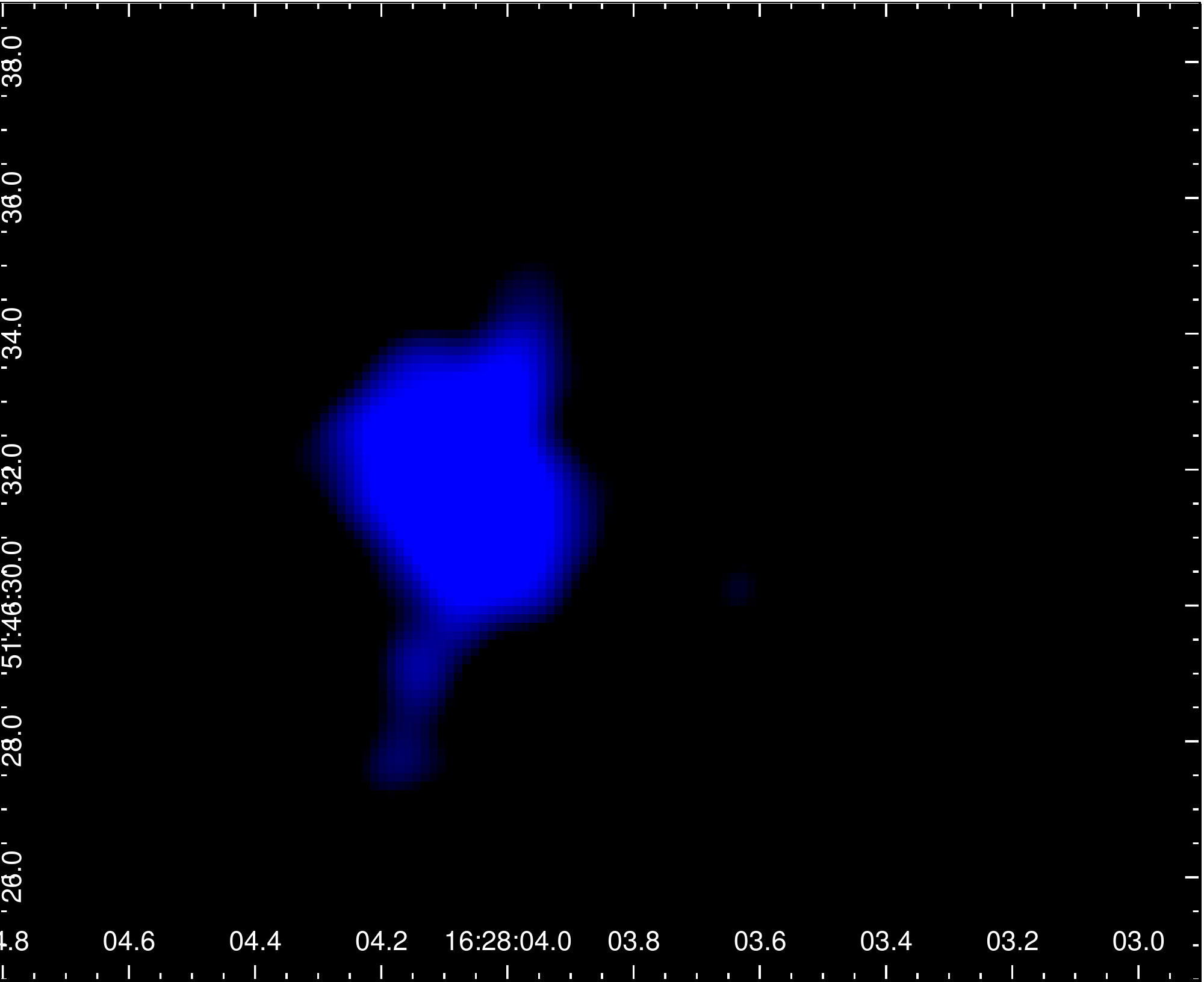}

\includegraphics[width=\textwidth,trim={5.2cm 0 2.2cm 0},clip]{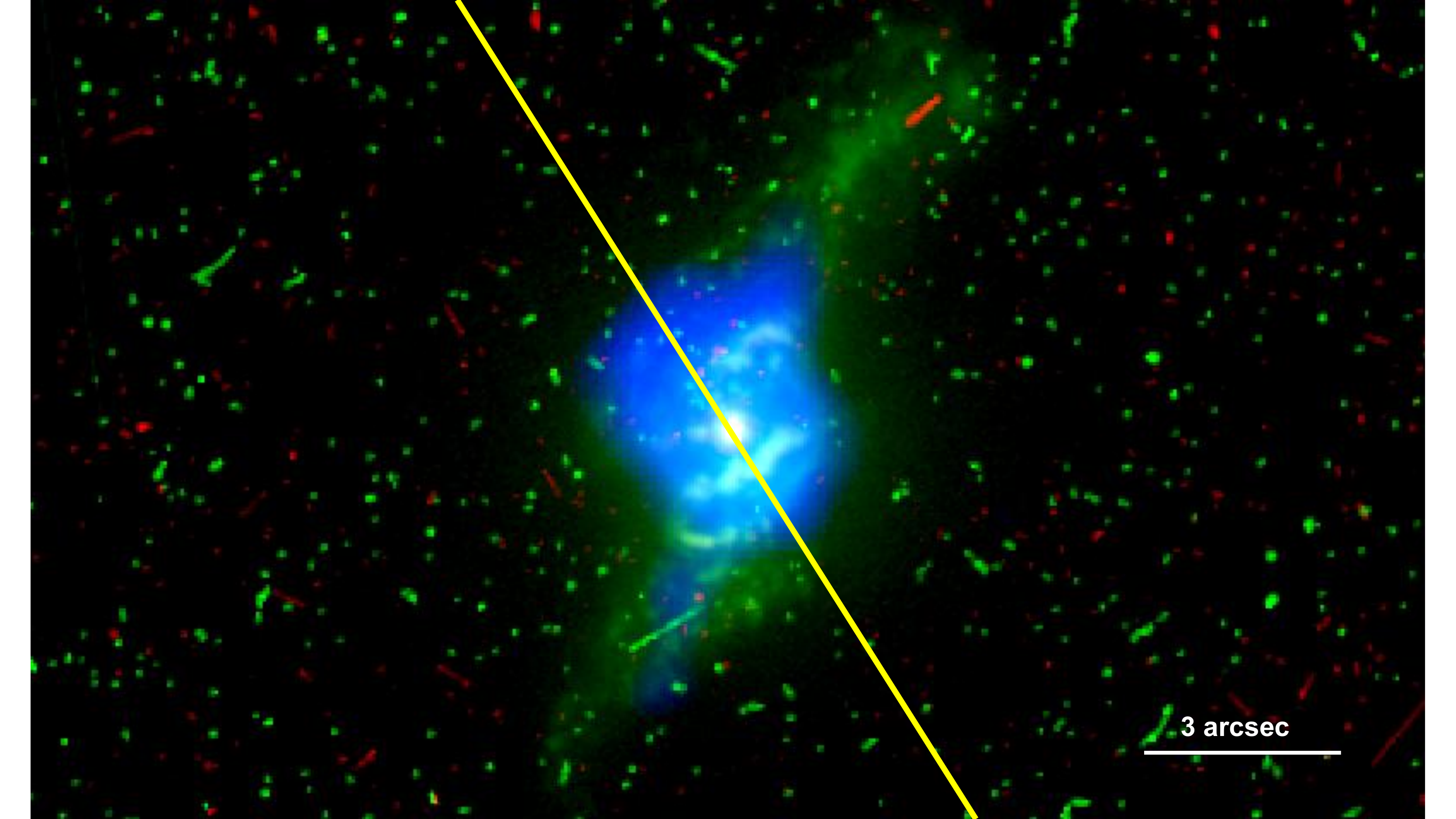}
\caption{(Upper-Left): H$\alpha$ image from the \emph{HST}, (Upper-Middle panel): [O${III}$] image from the \emph{HST}, (Upper-Right): soft X-ray from the \emph{Chandra} observation, and (Bottom):Composite image of Mrk\,1498 using H$\alpha$  emission line (red), [OIII] emission line (green), and soft X-rays in the 0.5-2 keV energy band (blue). The optical data were taken from the \emph{HST} archive and were processed using the sharp dividing method that allows to see the internal parts of the galaxies \citep{marquez1996}. In the images the gray levels
extend from twice the value of the background dispersion to the
maximum value at the center of each galaxy. We used IRAF to estimate these values, and (Right): same image but with the direction of the jet from the NVSS data in yellow.}
\label{indivlambda}
\end{figure*}

We used \emph{HST} data for imaging because of its high spatial resolution. These data have been processed following the sharp dividing method to show the internal structure of the galaxies \citep{marquez1996}. The gray levels extend from twice the value of the background
dispersion to the maximum value at the center of the galaxy.
We used the {\sc imexam} task in IRAF to estimate these values. The upper part of Figure \ref{indivlambda} shows the H$\alpha$ (left) and [OIII] (middle) emission line images. Two different extended emissions can be observed in addition to the nuclear emission. The circumnuclear extended emission extends to about 1 kpc and is observed in both images, whereas the large-scale extended emission is fainter and extends to about 7 kpc and is specially evident in the [OIII] image, as already reported in \citet{keel2015}. A multiwavelength comparison of the images is presented in Section \ref{multiimages}.

\subsection{X-rays/UV}

\begin{figure*}
\includegraphics[width=0.8\textwidth]{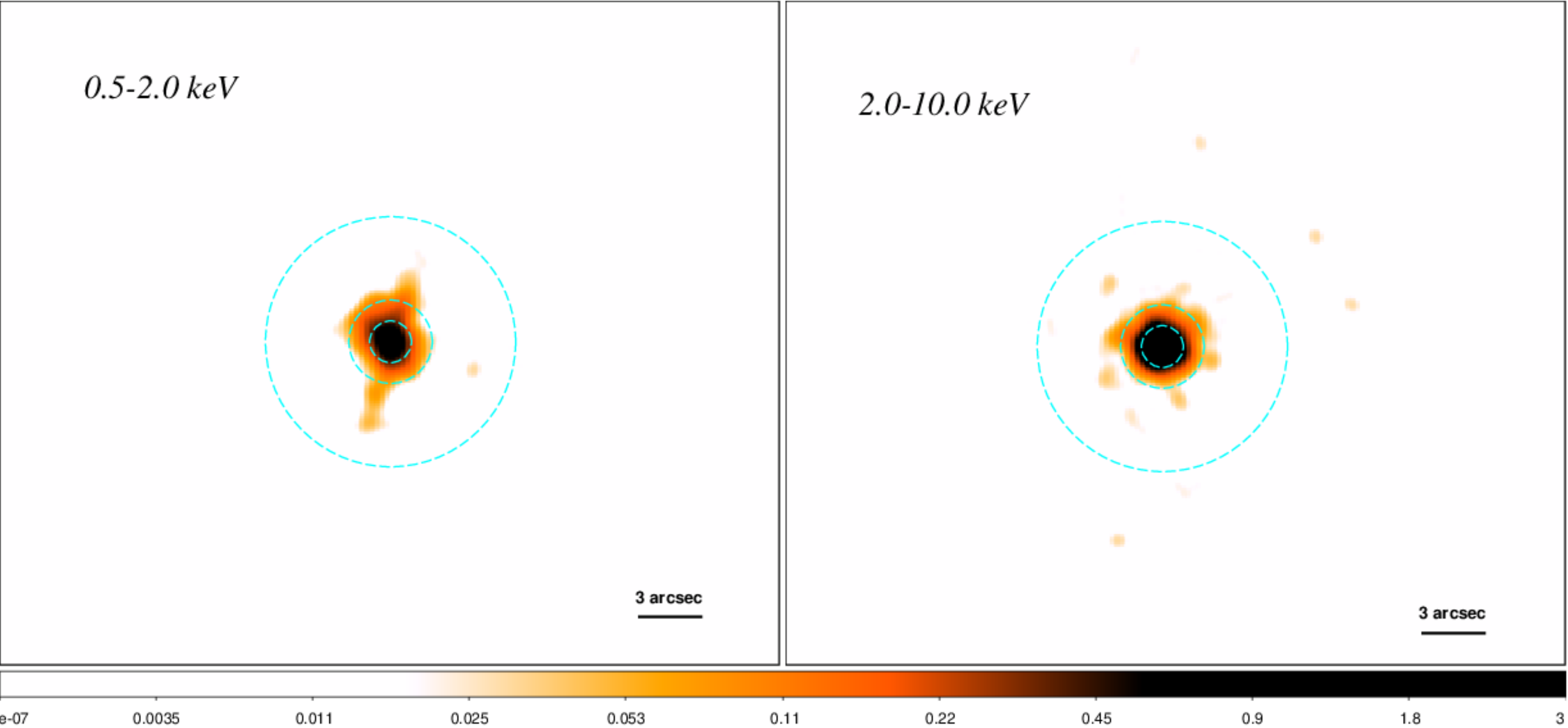}
\caption{Smoothed \emph{Chandra} images in the 0.5-2 keV (left) and 2-10 keV (right) energy bands. The color map shows the brightness levels of the source, which are matched for the two energy bands. The blue dashed lines represent circles with radius 1 arcsec, 2 arcsec, and 6 arcsec centred on the position of Mark\,1498.} 
\label{chandraimage}
\end{figure*}

The main result from the \emph{Chandra} imaging is that extended emission is observed at energies below 2 keV for the first time in this source, while a point-like source is observed in the 2-10 keV energy band, as shown in Figure \ref{chandraimage}. 
We have done a spatial analysis by comparing the fluxes of the unresolved and the extended emission. For this purpose we have extracted spectra from different spatial regions of the observed emission, one with a circular region of 6 arcsec radius, a second with an annular region between 1 arcsec\footnote{The fractional encircled energy at a radius of 1 arcsec encircles more than 80\% of the photons for an on-axis point source. http://cxc.harvard.edu/proposer/POG/html/chap4.html.} and 6 arcsec, and a third one from 2 arcsec (corresponding to the unresolved emission observed in the Chandra image) to 6 arcsec (the regions covered by these radii are represented as blue dashed lines in Figure \ref{chandraimage}). We find that the annular region with radii 1--6 arcsec is dominated by extranuclear emission both in the soft and hard energy bands. The extended emission above 2 arcsec contributes by a 5\% to the hard energy band and to a 37\% of the soft energy band. Furthermore, the emission between 2-6 arcsec reaches only 5 keV. This indicates that the resolved emission contributes only to soft X-ray energies by a factor of 37\% of the observed emission.

The X-ray data were taken from different satellites, thus the spectra were extracted with different apertures, and variability is one of the properties characterizing AGN \citep{peterson1997}. Therefore we analized the X-ray, UV, and optical variations in a timescale of 11 years using public available data from \emph{Swift}, \emph{Chandra} and \emph{XMM--Newton}. 

\begin{figure}
\includegraphics[width=0.5\textwidth]{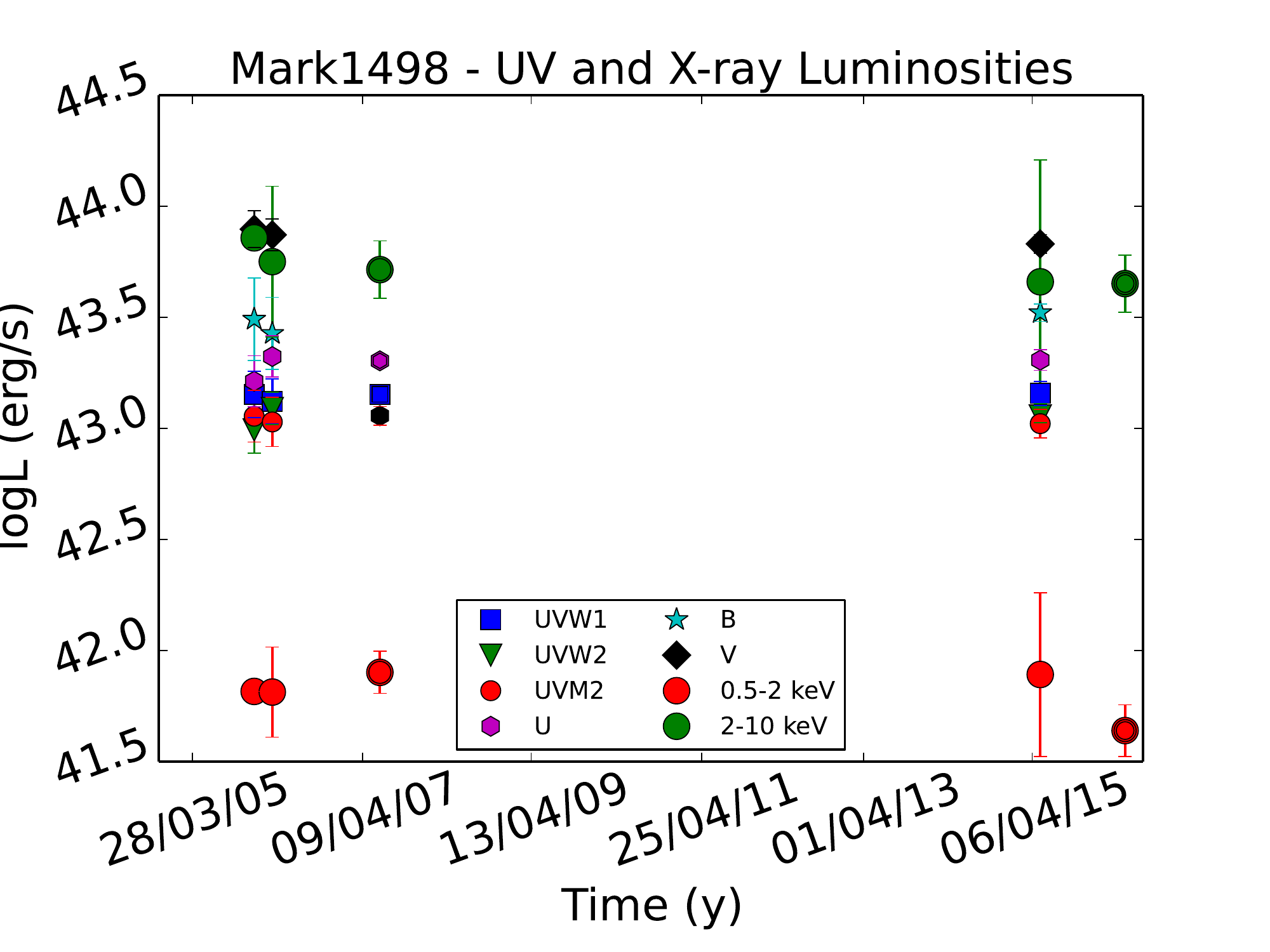}
\caption{Luminosities at X-rays in the 0.5-2 keV and 2-10 keV energy bands from \emph{Swift}/XRT (in 2005, 2006, and 2015), \emph{Chandra}/ACIS-S (in 2016), and \emph{XMM--Newton}/pn (in 2007), and in the optical/UV from \emph{Swift}/UVOT and \emph{XMM--Newton}/OM simultaneously with the X-ray observations. For comparison purposes, the model used for the spectral fit is composed by two power laws, one of them absorbed. }
\label{variability}
\end{figure}

We fitted the spectra with a simple model consisting on a power law representing the 0.5-2 keV energy band, plus an absorbed power law representing the 2-10 keV band. This simplistic model gave acceptable spectral fits to all the spectra, with a reduced $\chi^2$ in the range [0.8-1.3], therefore useful to estimate the luminosities of the source at different dates. We recall that optical/UV photometry were obtained simultaneously to X-ray data with \emph{XMM--Newton} and \emph{Swift} as explained in Section \ref{optmon}. The observed luminosities are reported in Figure \ref{variability}. As it can be seen, variations at these timescales are not observed at any energy range, with all the measurements being compatible within the errors. It is worth noticing that the X-ray luminosity in the 2-10 keV energy band agrees well with measurements obtained with \emph{Suzaku} in 2006 (logL(2-10 keV)=43.7) by \cite{eguchi2009} and \cite{kawamuro2016}.

These results indicate that the observed emission from Mrk\,1498 is not strongly variable in flux, thus we analyzed simultaneously the observations taken from different instruments. The \emph{XMM--Newton}/pn, \emph{NuSTAR} and \emph{Swift}/BAT data have also been analyzed in \cite{ursini2018}, who found the nucleus to be absorbed by a column density of $\sim 2 \times 10^{23}$ cm$^{-2}$. The primary emission is well described by a power law with a photon index of $1.5 \pm 0.1$ with a high-energy cut-off of $80^{+50}_{-20}$ keV \citep{ursini2018}.  We re-analyzed the \emph{XMM--Newton}/pn, \emph{NuSTAR} and \emph{Swift}/BAT data jointly with the \emph{Chandra} and \emph{Swift}/XRT data.

On the assumption of an obscured AGN in the nucleus of Mrk\,1498, we performed broad-band (0.5--195 keV) fits testing a physical absorption model by a torus. We used the \textsc{borus} model, which describes absorption with self-consistent Compton reflection and iron fluorescent lines from a smooth gas torus \citep{balokovic2018}. In this model the half-opening angle, $\sigma$, and the inclination angle, $i$, of the torus are free parameters, thus we can use the results obtained from the mid-infrared using \emph{Spitzer} to constrain the physical properties (see Table \ref{tab:midinfraredfit}). We used the values of the smooth model because this is the distribution assumed by \textsc{borus}. It is worth noting that the angles are the complementary of those obtained from the \emph{Spitzer} data. We included the scattered and line emission components, linking their column densities to that of the absorber (``coupled'' model).
The primary continuum was modelled with a cut-off power law. 
The photon index, $\Gamma$, and normalization, $norm$, of both the scattered and line components were tied to those of the primary power law, allowing for a free relative normalization by means of a multiplicative constant. 
As already discussed in \cite{ursini2018}, the spectrum shows a soft excess below 1 keV, on top of the absorbed power law. 
Following their work, we included both a secondary, unabsorbed power law to model the scattered component (with a free scattered fraction $f_s$), and the \textsc{mekal} model, representing thermal plasma either photoionized or due to a starburst component  \citep{bianchi2010, lore2017b}. 
We kept all the parameters tied between the different observations, adding a free multiplicative constant to account for cross-calibration issues. The constant was found to be in the range 0.9--1.1 with respect to \emph{NuSTAR}.

\begin{figure}
\includegraphics[width=0.5\textwidth]{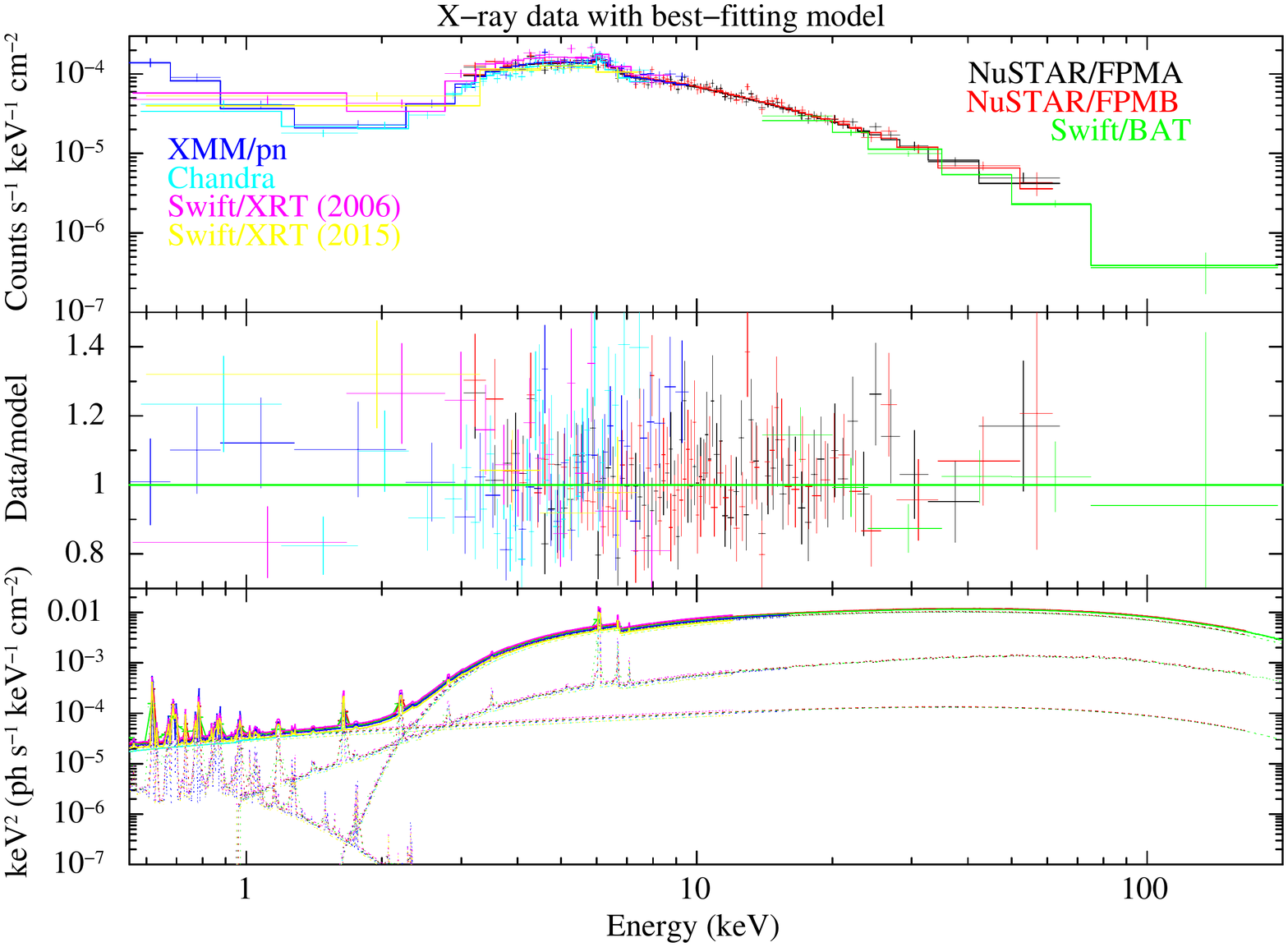}
\caption{Upper panel: X-ray spectrum from XMM/pn (blue), Chandra (cyan), 2006 Swift/XRT (magenta), 2015 Swift/XRT (yellow), NuSTAR/FPMA (black), NuSTAR/FPMB (red) and Swift/BAT (green), plotted with the best-fitting model. Middle panel: Data/model ratio. Lower panel: Bets-fitting \textsc{Borus} model including the unabsorbed power law and \textsc{mekal}. The data were binned for plotting purposes.}
\label{fig_x}
\end{figure}

\begin{figure}
\includegraphics[width=0.5\textwidth]{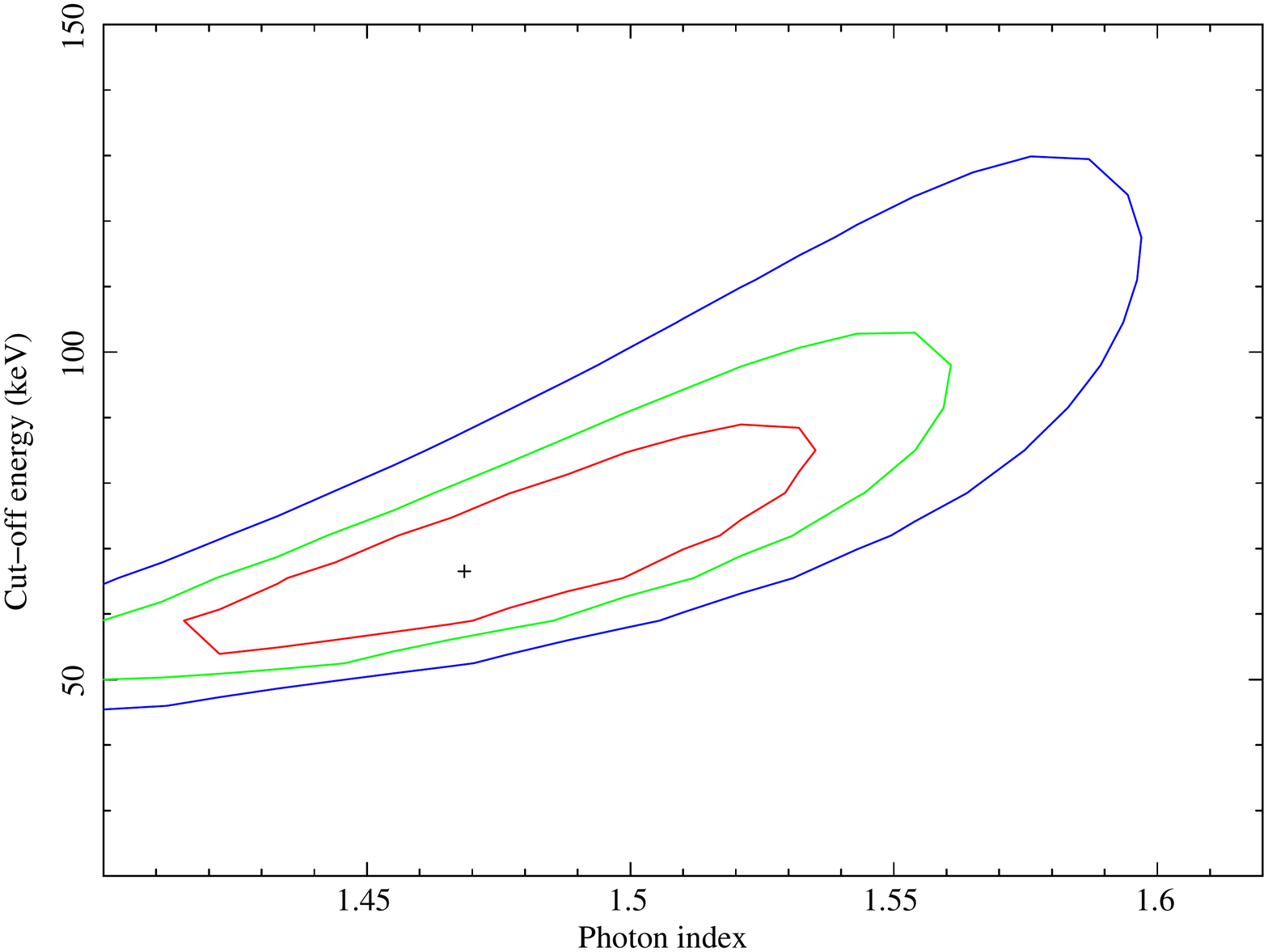}
\caption{Contour plot of the spectral index, $\Gamma$, versus the cut-off energy at 1$\sigma$, 2$\sigma$, and 3$\sigma$. Note that in the borus model $\Gamma >$ 1.4. }
\label{contours}
\end{figure}

The results are summarized in Table \ref{fit_x} and the spectral fit is presented in Figure \ref{fig_x}. We included a contour plot of $\Gamma$ and the cut-off energy in Figure \ref{contours}. Here it can be seen that, although there might be some degeneracy between these parameters, $\Gamma$ is smaller than 1.6 at a 3$\sigma$ confidence level. We note that our results are in agreement with those in \citet{ursini2018} even if different models have been used for the data analysis.

 \begin{table}
 \begin{center} 
  \caption{ \label{fit_x}X-ray best-fitting parameters. }
 \begin{tabular}{  c c c  } 
 \hline
Borus & log$N_H$ (cm$^{-2}$) & 23.22$\pm$0.02 
 \\ cutoffpl & $\Gamma^a$ & 1.47$\pm$0.07
 \\ cutoffpl & Cut-off (keV) & 67$^{+26}_{-11}$ 
 \\  cutoffpl & norm (10$^{-3}$) & 3.0$^{+0.5}_{-0.2}$ 
 \\  mekal & kT (keV) & 0.26$^{+0.09}_{-0.05}$ 
 \\  mekal & norm (10$^{-5}$) & 2.5$\pm$0.9 
 \\  constant & $f_s$ (10$^{-2}$) & 1.1$\pm$0.2 
 \\  L(2-10 keV) & (erg \hspace{0.1cm} s$^{-1}$) & (0.9--1.3)$\times$10$^{44}$
 \\  
 \hline
 \end{tabular} \end{center} 
		\begin{flushleft}
			\textit{Note.} $^a$ The photon index in \textsc{borus} cannot be lower than 1.4.
		\end{flushleft}
 \end{table} 
 
\subsection{\label{multiimages} Multiwavelength imaging}

Data taken from the \emph{HST}, \emph{Chandra} and VLA can be observed in Figure \ref{indivlambda}. We first aligned the images by using the peak of the nuclear source as the reference point. We used ds9\footnote{http://ds9.si.edu/site/Home.html} to determine this maximum and also to perform a combined color image to compare the different structures. The red data refers to the H$\alpha$ emission line, the green to the [OIII] emission line, the blue to the soft X-ray emission, and the yellow line represents the direction of the jet in the NVSS image. This image shows that the rotation axis of the circumnuclear extended [OIII] emission line (at a distance of about 1 kpc) and soft X-ray structures are aligned with the large-scale radio jet. The large-scale extended emission (at a distance of about 7 kpc), observable in [OIII], however, is not aligned with the radio jet.


\section{Discussion}

The multiwavelenth analysis presented in this paper for the nucleus in Mrk\,1498 shows emission from different structures at different spatial scales. Here we report the main results that will be discussed in the following:

\begin{itemize}
    \item The presence of the lobes in this GRG indicate that this emission is very old, contrary to the GPS detected in the very center that indicates a young nucleus.
    \item Mid-infrared frequencies are dominated by the torus emissions. The \emph{Spitzer} spectrum shows that the torus is observed edge-on with a relatively small covering fraction.
    \item The large-scale jet observed in the radio and the extended emission observed by \emph{HST} in the [OIII] emission line are not aligned. This is unusual because when AGN photoionization is reponsible for the emission in the NLR these structures are aligned in the nuclear regions.
    \item Emission lines in the optical spectrum are fitted by three Gaussian components, narrow, intermediate and broad. Our classification of a Seyfert 1.8 is compatible with the edge-on view obtained in the mid-infrared.
    \item The X-ray spectra are well reproduced by an intrinsic plus a reprocessed components. We used the configuration obtained at mid-infrared frequencies to fit the data, showing that the nucleus is highly absorbed.
\end{itemize}

\subsection{\label{disc:AGN}The AGN in Mrk\,1498}

The nature of the AGN in the nucleus of Mrk\,1498 is controversial. On the one hand, the X-ray data obtained by \emph{Suzaku} was used by \cite{eguchi2009} to show that the nucleus is highly obscured by a column density of (1.3$\pm$0.4)$\times$10$^{23}$cm$^{-2}$, as well as by \cite{ursini2018}, who fitted \emph{XMM--Newton}+\emph{NuSTAR} data and reported an $N_H \sim$2$\times$10$^{23}$cm$^{-2}$. On the other hand, \cite{keel2015} used optical \emph{HST} data and found extended emission around the nucleus that they explained as clouds ionized by a faded AGN. 

In this work we have used multiwavelength data at different spatial scales to study this AGN.
The radio image from the \emph{VLBA} shows an unresolved source with an elongation towards the NE/SW. At this mas scales, we are detecting the most recent emission due to the accelerated plasma in the vicinity of the central black hole, therefore giving a strong indication that the AGN is active. Moreover, the radio spectrum shows the presence of a GPS, indicative of a young AGN \citep{bruni2019}. The X-ray image from \emph{Chandra} also shows a point-like source with extended emission observable at soft energies.
Furthermore, the \emph{Spitzer} spectrum shows the typical shape of an AGN dominated by the torus emission \citep{omaira2017}. All these data strongly support the idea that an active nucleus resides in the nucleus of Mrk\,1498 at the present epoch. 

In order to test whether the nucleus is able to produce the ionization of the observed nuclear emission or not, we compare the bolometric luminosity of the source with the ionizing luminosity reported in \cite{keel2012} of $L_{ion} > 3.1 \times 10^{43} erg \hspace{0.1cm} s^{-1}$ using a long slit spectrum from the Lick Telescope. From the X-ray data, the luminosity in the 2--10 keV energy band is $\sim 10^{44} erg \hspace{0.1cm} s^{-1}$, so with the bolometric correction in \cite{runnoe2012} we estimate $L_{bol} = 23.31L (2-10 keV)  \sim 10^{45} erg \hspace{0.1cm} s^{-1}$, suggesting that the AGN is capable enough to power all the ionizing emission that we observe in the nuclear region, showing conclusive evidence of an obscured but active nucleus, and excluding the case of a fading AGN.

The X-ray spectra are indeed modelled by a power-law emission in the 2--195 keV energy band, and this emission is highly absorbed by a column density of 1.7$\pm$0.1$\times$10$^{23}$cm$^{-2}$, in good agreement with \cite{eguchi2009} and \cite{ursini2018}. If we compare the column density with those of other optically classified Seyfert 1.8/1.9 sources \citep{lore2017a}, only two out of the fifteen studied sources have $N_H>$10$^{23}$cm$^{-2}$, thus Mrk\,1498 is among the most obscured Seyfert 1.8/1.9, in the range of Seyfert 2s \citep{lore2015}. 
The X-ray absorption is usually interpreted as a measure of the absorption by the torus, with Seyfert 2s thought to be observed directly throughout the torus. 
This absorption can be compared with the optical depth derived from mid-infrared observations, that also trace signatures of the dust emission. Assuming a constant gas-to-dust ratio and using the relation between the optical depth and column density \citep{bohlin1978,tang2016}, we obtain $N_H \sim $ 1.8$\times$10$^{23}$cm$^{-2}$ for the clumpy model, and $N_H \sim $ 2.0$\times$10$^{22}$cm$^{-2}$ for the smooth model, both in agreement with the X-ray absorption, indicating that the media responsible for the emission might be connected. We notice that the inferred $N_H$ from the mid-infrared data represents the measurement of the absorption from the equator of the torus, whereas our line of sight is about 30-40 degrees from above the equator, thus the $N_H$ represents an upper limit. 
The \emph{Spitzer} data at mid-infrared frequencies allowed us to determine the geometry of the torus, which shows that it has a small covering fraction of $f_{cov}=0.40\pm0.02$ and that we are observing from above its edge. This is consistent with the Seyfert 1.8 classification obtained in the optical, as expected from unification schemes, where orientation is responsible for the different AGN classification. Seyfert 1.8 are thought to be observed at 30-40 degrees above the equator of the torus, our line of sight partially intercepts the torus but the BLR is still visible \citep{antonucci1993}. The fact that slight silicates are observed in emission at 10-18 $\mu$m also favours the type 1 classification, because this means that we are observing the source edge-on.  
Using the \emph{Spitzer} data we can also estimate the size of the torus from its width, defined as $Y = R_{out}/R_{in}$, where $R_{in}=0.4 \sqrt{L_{bol}/10^{45}}$ [pc]. From this and using the results given in Table \ref{tab:midinfraredfit} we obtain a size of $\sim$ 12 pc in the case of a clumpy torus and of $\sim$ 6 pc for a smooth distribution. These values are in agreement with those obtained
for nearby Seyfert galaxies (Gonzalez-Martin et al., submitted).
Thus the large observed absorption might be explained by absorption coming from a multiphase gas originated at different regions instead of by the torus alone. \cite{wada2015} explained AGN obscuration as circulation of gas driven by the central radiation source in vertical forms, resulting in outflows, which naturally produces a toroidal geometry. In fact, their numerical simulations showed that for luminosities of the order of 10$^{44}$ erg \hspace{0.1cm} s$^{-1}$, the covering fraction is smaller than 0.5, in agreement with our results. We also inferred a larger $N_H$ from the mid-infrared data than that obtained directly from the X-rays, consistent with a lower amount of gas compared to that estimated from the dust emission.
Furthermore, one of the characteristics of GPS sources is the presence of dense gas in their environments, that could also contribute to the large amount of obscuration observed at X-rays.
Assuming the presence of a GPS in the very center of the nucleus of Mrk\,1498, we can estimate its projected linear size from the turnover frequency \citep{odea1998}. In this case we obtain a size of about 0.1 kpc using NVSS and Effelsberg data, roughly corresponding to the size of the plasma linked to the new radio phase, that includes the newborn jets. From the VLBA map at milliarcsec resolution, we obtained a projected linear size lower than 1.5 pc for the central radio core. 
Thus we confirm the obscured AGN nature of the active nucleus in Mrk\,1498, where the absorption might not come only from the torus but also from other components in our line of sight.


\subsection{\label{restarting}Restarting activity}

The Effelsberg data presented in \cite{bruni2019} shows a GPS spectrum in the nucleus of Mrk\,1498. 
Historically, the nature of GPS has been explained as a core embedded in a dense environment, in which it fails to expand, resulting in a compact source with an absorbed spectrum \citep{vanbreugel1984,bicknell1997,marr2014,tingay2015};
or as a self-absorbed source due to a very dense plasma, because the source is young so did not have time to expand, and thus the low frequencies are self-absorbed from the same plasma that emits it \citep{snellen2000,fanti2009}.
To date, observations favored the second scenario and
GPS are considered as the best representation of young radio sources, i.e., with ages in the range of 10$^2$-10$^5$ yr \citep{fanti1995,perucho2016,bruni2019}. 
Conversely, the large scale radio image of this galaxy shows a GRG with a size of 1.2 Mpc. The lobes of these galaxies are thought to be as aged as 10$^8$ yr \citep{alexander1987}, thus there exists a discrepancy within these timescales because the new born jets cannot be responsible for the feeding of the old lobes.
The most natural explanation to solve this would be that the nucleus in Mrk\,1498 has restarted its nuclear activity, that is thought to occur every 10$^4$-10$^8$ yr \citep{reynolds1997}. The new nucleus shows elongated emission in the same direction as the old jets. This reminds to typical double-double radio galaxies \cite[DDRG, e.g.,][]{schoenmakers2000}, whose morphology is characterized by two pairs of radio lobes aligned along the same direction and having a common nucleus. In the particular case of Mrk\,1498 this morphology is not observed, but the mas scale VLBA data shows elongated emission in the same direction as the lobes observed in the VLA image, suggesting that Mrk\,1498 may be the progenitor of a DDRG.

This result might indicate that we are missing a fraction of AGN where the nuclear activity has been restarted because of selection effects. Indeed, while the percentage of galaxies showing linear sizes larger than 0.7 Mpc is very low, with only a 6\% of them in the 3 CR Catalogue \citep{ishwara1999}, \cite{bassani2016} undertook a radio/X-ray study of the combined \emph{INTEGRAL}+\emph{Swift} AGN populations, and
found that a 22\% of their sample are GRG, an impressive number compared to \cite{ishwara1999}. Remarkably, 61\% of these sources show signs of restarting
activity, as a presence of a GPS is observed from Effelsberg data \citep{bruni2019}. Mrk\,1498 is among these sources, with its GPS peaking at 4.9 GHz (see Figure \ref{Eff}). Thus we emphasize the observational result reported by \cite{bassani2016} that hard X-rays could be a tool to find GRG galaxies with restarted activity.


\subsection{\label{extended}The extranuclear emission}

Mrk\,1498 shows two remarkable features especially evident in the \emph{HST} image (Figure \ref{indivlambda}). At 0.6 - 1.8 kpc scales, multiple circumnuclear rings of ionized gas ([OIII]$\lambda \lambda$4959,5007 and H$\alpha \lambda$6563) are visible, possibly tracing the ionization cones of the AGN \citep{keel2015}. In fact, based on integral field spectroscopy (IFS) observations with GEMINI/GMOS,  \cite{keel2017} have shown that the circumnuclear rings follow the main rotation of the galaxy. In addition, a match between the [OIII] and the 0.5-2 keV emission from \emph{Chandra} in the inner 2 arcsecs (see Figure \ref{chandraimage}), possibly suggests a narrow line region origin \citep[e.g.,][]{bianchi2006,gomezguijarro2017}. In radio galaxies, these structures can also be aligned with the radio jet \citep{balmaverde2012}, as seems to be the case of Mrk\,1498, where the elongation of the soft X-ray and [OIII] emission line regions are aligned in the same direction as the jet. 

On the other hand, relatively large-scale structures extended out to 10 arcsec ($\sim$10 kpc) to the northeast and southwest are present and particularly bright in [OIII]. 
Interestingly, this large-scale structure is not aligned neither with the radio jet axis (observed with the VLA, see Figure \ref{indivlambda}) nor with the soft X-ray emission, making its origin intriguing. We discuss some of the possible scenarios.

The misalignment and the large scale extension of such diffuse [OIII] emission do not support a NLR origin \citep{husemann2014}. In order to confirm this statement, we run grids of photo-ionization models with CLOUDY \citep[version C17.01][]{ferland2017} to investigate if the photo-ionization by the central nuclear source alone is able to ionize the clouds at a distance of d = 7 kpc from the nucleus. 
The input parameters are the gas density, which we allow to vary between 1 cm$^{-3}$ and 10$^4$ cm$^{-3}$, the distance of the clouds to the nucleus,
the spectral energy distribution of the ionizing radiation 
\citep{mathews1987}, solar abundances and clouds with no dust, and the bolometric luminosity of log(L$_{bol}$)=45 (see Sect. \ref{disc:AGN}). We compared the results of optical line ratios in the simulation with the values reported for a spectrum of the northern extra-nuclear emission reported in \cite{keel2012}. The simulation resulted in much smaller values than those reported in \cite{keel2012}, except for densities below 10 cm$^{-3}$, suggesting that the nuclear luminosity of Mrk 1498 is not able to sustain extended emission at such large distances. 
For comparison, we performed the same simulation for the Teacup galaxy, which is also included in the sample of \cite{keel2012} and has been confirmed as a faded AGN. For this galaxy, log(L$_{bol}$)=45 and d=11.3 kpc. The values of the line ratios in the simulation agree well within errors with the observational results for a gas density in the range between 10-100 cm$^{-3}$, in agreement with the faded AGN scenario in the case of the Teacup galaxy. 
 However, we note the presence of an intermediate component in the fitting of the optical nuclear spectrum, which may suggest the presence of non-rotational motions. This broad component ($\sigma$ = 485 km$\hspace{0.1cm}$s$^{-1}$) is slightly shifted respect to the systemic velocity, thus a possible interpretation is that it probes the bulk of an outflow \citep{fiore2017}. Eventually, we cannot discard the possibility of outflowing material in the NLR, but the kinematics of the large-scale extra-nuclear emission need further investigation.

The interaction between a jet and the inter stellar medium can also produce extended emission in form of outflows \citep{best2000,lore2018}. The misalignment between the emission line regions, the soft X-ray emission and the jet makes this possibility unlikely. Nevertheless, a recent work by \cite{balmaverde2019} shows that in FRII radio galaxies the optical line emission structures are preferentially oriented perpendicularly to the radio jets, suggesting that we are most likely observing gas structure associated with the secular fueling of the central SMBH. Further research needs to be done in order to understand if the emission by the radio jets are related to the optical emission line structures.

\cite{keel2015} used the \emph{HST} imaging to subtract the host galaxy contribution to the AGN. 
This technique is usually used in the study of galaxy clusters, where the galaxies can be relaxed or present indications of distorted structures \citep{guennou2014,martinet2016}.
Mrk\,1498 showed two broad spiral bands that persist after the subtraction, so they suggested that this galaxy might be the aftermath of a merging event. 
Recently, \cite{blecha2018} studied the role of galaxy mergers in fuelling AGN through the Wide-field Infrared Survey Explorer (WISE). They proposed a two-colour diagnostic to select merging AGN with W1 - W2 $>$ 0.8, and W2 - W3 $>$ 2.2. We searched for the colours of Mrk\,1498, and found W1 - W2 $=$ 1.03, and W2 - W3 $=$ 2.95, which locates this galaxy in the late-post merger phase, i.e., 10-150 Myr after the merging occurred. 
The optical spectrum of Mark\,1498 shows that the galaxy might have experienced a recent burst of star formation about 10$^6$ years ago, thus this could also support the post merger scenario, since galaxies in the post starburst phase usually show signs of young stellar populations \citep{pracy2014}. 
However, the fact that the large and small scale radio jets are aligned makes unclear the hypothesis that a major merger occurred in Mrk\,1498. 
We recall that secular processes such as minor interactions within the galaxy can also produce distorted structures. As shown by \cite{treister2012}, in 90\% of AGN a triggering may be activated after secular processes, such as minor interactions, while only in the most luminous AGN (10\%) major mergers reactivated the nuclear activity.
We note that the restarting activity of the nucleus proposed here would naturally explain the triggering of the nuclear activity and the production of the observed tidal streams \citep{stockton1983}.
The interpretation in \cite{keel2017},
although possible, is severely hindered by the limited field of view (FoV) of their IFS observations covering only the innermost region of 3.5 arcsec $\times$ 3.5 arcsec. 

A spectrum of the northern extended region was presented in \cite{keel2012}. Interestingly, the line intensity ratios of this emission is compatible with emission from HII regions using standard diagnostic diagrams \citep[e.g.,][]{baldwin1981}. This emission could therefore be gas ionized by young massive stars. Furthermore, our SFH obtained from the stellar populations analysis shows that the galaxy might have experienced a recent burst ($\sim10^6$ years old) of star formation (see Figure \ref{sfh}), that correspond to the current 0.01-0.1\% of stellar mass of the galaxy, in well agreement with the triggering and restarting activity scenario (see Sect. \ref{restarting}). 

Hence, a mapping of the kinematics of the [OIII] bright diffuse emission with integral fiel spectroscopy (IFS) data in Mrk\,1498 is required to shed light on its intriguing origin and to differentiate between the various scenarios proposed here.


\section{Conclusions}

In this work we have used multiwavelength data at radio, mid-infrared, optical, ultraviolet, and X-ray frequencies to study the innermost parts of the galaxy Mrk\,1498. We find that:

\begin{itemize}
    \item At radio frequencies, it is a GRG with a linear size of 1.2 Mpc estimated from the NVSS image at 1.4 GHz. The \emph{VLBA} image at pc-scale shows an almost unresolved ($<$1.5 pc) central component with extended emission towards NE/SW in the same direction as the largest scale image. The radio spectrum of its core is fitted by a log-parabola model with a peak at 4.9 GHz, indicative of a young radio phase, implying the reactivation of the nuclear activity.
    \item The \emph{Spitzer}/IRS data at mid-infrared frequencies shows the typical shape of an AGN dominated by the torus emission. The geometry of the torus is characterized by a viewing angle of $\sim$ 30 degrees, and an opening angle in between of 15-30 degrees, both measured from the equatorial plane.
    \item The optical data obtained from the San Pedro M\'artir Telescope shows a spectrum of a Seyfert 1.8. We fit the H$\beta$-[OIII] region with a three-components model with narrow, intermediate and broad widths. From the SFH we find that the galaxy might have experienced a burst of star formation about $10^6$ years ago. 
    \item The X-ray/UV data does not show signs of variability in a timescale of 11 years, thus all the spectra were fitted together. We fitted the data using a model of an obscured AGN using the parameters of the torus obtained with the \emph{Spitzer} data, and obtained a column density of $N_H = (1.7 \pm 0.08) \times 10^{23} cm^{-2}$. For the first time, we report extended emission in the 0.5--2 keV energy band from the \emph{Chandra} image.
\end{itemize}

Taking into account the results presented here we confirm the presence of an obscured but active AGN in the center of Mrk\,1498, excluding the fading AGN scenario. We propose that this galaxy is the result of a galaxy merger or secular processes, such as minor interactions, responsible for the reactivation of the nuclear activity. This event that reactivated the nuclear activity, gave place to the actual young nucleus, that resembles to be the progenitor of a DDRG. At the high spatial resolution of \emph{HST}, it shows extended emission in the [OIII] emission line which does not seem to be related to phototoionization from the AGN but could be either explained by star formation or outflowing material. IFS data are required to confirm our interpretation.


\section*{Acknowledgments}

We are grateful to Dr. J. Masegosa and Dr. I. M\'arquez for helpful discussions during this work. We thank the anonymous referee for her/his comments that helped to improve the manuscript. 
L.H.G. acknowledges financial support from FONDECYT through grant 3170527.
V.C. acknowledges support from CONACyT research grant 280789. F.P., L.B., G.B. acknowledge ASI-INAF agreement I/037/12/0. G.B. acknowledges financial support under the INTEGRAL ASI-INAF agreement 2013-025-R1.
S.C. acknowledges AYA 2016-76682-C3-1-P and from the State
Agency for Research of the Spanish MCIU through the "Center of Excellence Severo
Ochoa" award for the Instituto de Astrof\'isica de Andaluc\'ia (SEV-2017-0709)”.
P.A. and Y.D. acknowledge financial support from Conicyt PIA ACT 172033.
E.F.J.A  acknowledges support of the Collaborative Research Center 956, subproject A1, funded by the Deutsche Forschungsgemeinschaft (DFG). 
O.G.M acknowledges financial support from UNAM PAPIIT IA 103118.

This paper is based on observations obtained with {\it XMM--Newton}, ESA science mission with instruments and contributions directly funded by ESA Member States and the USA (NASA).
This research has made use of the
NuSTAR Data Analysis Software (NuSTARDAS) jointly developed by the ASI Science Data Center (ASDC, Italy) and the California Institute of Technology (USA). 
This work made use of
data supplied by the UK Swift Science Data Centre at the University of Leicester.
The scientific results reported in this article are based on data obtained from the Chandra Data Archive, observations made by the Chandra X-ray Observatory.
This work is based in part on observations made with the Spitzer Space Telescope, which is operated by the Jet Propulsion Laboratory, California Institute of Technology under a contract with NASA.
The National Radio Astronomy Observatory is a facility of the National Science Foundation operated under cooperative agreement by Associated Universities, Inc. 
Partly based on observations with the 100-m telescope of the MPIfR (Max-Planck-Institut für Radioastronomie) at Effelsberg.
This study is based on observations from the Observatorio Astron\'omico Nacional at San Pedro M\'artir (OAN-SMP), Baja California, M\'exico.
Based on observations made with the NASA/ESA Hubble Space Telescope, and obtained from the Hubble Legacy Archive, which is a collaboration between the Space Telescope Science Institute (STScI/NASA), the Space Telescope European Coordinating Facility (ST-ECF/ESA) and the Canadian Astronomy Data Centre (CADC/NRC/CSA).
This research has made use of NASA's Astrophysics Data System and of data, software and web tools obtained from NASA's High Energy Astrophysics Science Archive Research Center (HEASARC), a service of Goddard Space Flight Center and the Smithsonian Astrophysical Observatory,  the NASA/IPAC extragalactic database (NED), the STARLIGHT code, and the IRAF software.


\bibliographystyle{mn2e}
\bibliography{000referencias}


\end{document}